\documentclass[11pt]{article}

\usepackage{import}
\usepackage{BasicPlusMatrix}

\begin{document}

\title{Rigidity in Mechanism Design and its Applications}

\author{Shahar Dobzinski \and Ariel Shaulker\thanks{Weizmann Institute of Science.  Emails: {\ttfamily\{shahar.dobzinski, ariel.shaulker\}@weizmann.ac.il}. Work supported by ISF grant 2185/19 and BSF-NSF grant (BSF number: 2021655, NSF number: 2127781).}}

\maketitle

\begin{abstract}
We introduce the notion of rigidity in auction design and use it to analyze some fundamental aspects of mechanism design. We focus on the setting of a single-item auction where the values of the bidders are drawn from some (possibly correlated) distribution $\mathcal F$. Let $f$ be the allocation function of an optimal mechanism for $\mathcal F$. Informally, $S$ is (linearly) rigid in $\mathcal F$ if for every mechanism $M'$ with an allocation function $f'$
where $f$ and $f'$ agree on the allocation of at most $x$-fraction of the instances of $S$, it holds that the expected revenue of $M'$ is at most an $x$ fraction of the optimal revenue.

We start with using rigidity to explain the singular success of Cremer and McLean's auction assuming interim individual rationality. Recall that the revenue of Cremer and McLean's auction is the optimal welfare if the distribution obeys a certain ``full rank'' condition, but no analogous constructions are known if this condition does not hold. We show that the allocation function of the Cremer and McLean auction has logarithmic (in the size of the support) Kolmogorov complexity, whereas we use rigidity to show that there exist distributions that do not obey the full rank condition for which the Kolmogorov complexity of the allocation function of every mechanism that provides a constant approximation is almost linear.

We further investigate rigidity assuming different notions of individual rationality. Assuming ex-post individual rationality, if there exists a rigid set then the structure of the optimal mechanism is relatively simple: the player with the highest value ``usually'' wins the item and contributes most of the revenue. In contrast, assuming interim individual rationality, there are distributions with a rigid set $S$ where the optimal mechanism has no obvious allocation pattern (in the sense that its Kolmogorov complexity is high). Since the existence of rigid sets essentially implies that the hands of the designer are tied, our results help explain why we have little hope of developing good, simple and generic approximation mechanisms in the interim individual rationality world.
\end{abstract}

\thispagestyle{empty}
\newpage

    \setcounter{page}{1}

\section{Introduction}

\subsubsection*{The Setting}

We consider the following standard auction model: there is one item for sale and $n$ bidders. The value of bidder $i$ for getting the item is $v_i$, and $0$ otherwise. $v_i$ is the private information of each bidder $i$. The values are drawn from some joint distribution $\mathcal F\in \mathbb R^n$ that is publicly known. This paper aims to design (deterministic) dominant strategy mechanisms that maximize the revenue. The literature considers two main notions of individual rationality:

\begin{itemize}
\item \emph{Ex-post individual rationality}: the payment of bidder $i$ is at most $v_i$ if he wins the item, and $0$ otherwise.
c\item \emph{Interim individual rationality}: the expected payment of bidder $i$ with value $v_i$ is at most $x_i\cdot v_i$, where $x_i$ is the probability that bidder $i$ wins the item given that his value is $v_i$. Recall that the auction mechanism is deterministic, so the expectation is only over the distribution $\mathcal F$ of instances.
\end{itemize}

When the values of the players are drawn from independent distributions, Myerson  \cite{Myerson81} provides a characterization of the optimal auction. In a sharp contrast, when the distribution of values is correlated, no such crisp characterization is known. Broadly speaking, the literature takes two approaches: developing approximation mechanisms, and identifying special cases where the optimal solution takes a simple form. The first approach is dominant when assuming ex-post individual rationality, whereas the second one is more prominent when assuming interim individual rationality.

We now give a brief survey of the most related literature. We start with surveying the state-of-the-art in the design of ex-post IR mechanisms. The jewel in the crown here is Ronen's 2001 paper \cite{ronen-lookahead} that can be seen as an early precursor to many later trends in Algorithmic Mechanism Design. In that paper, Ronen presents the lookahead auction: the revenue-maximizing auction in which the item can be sold only to the bidder with the highest value. 

Remarkably, Ronen proves that this simple, easy-to-describe auction always guarantees half of the optimal revenue. In contrast, computing the optimal auction is NP-hard \cite{Papadimitriou-Pierrakos}. Subsequent work considered a natural extension of the lookahead auction -- the $k$-lookahead auction. This is the revenue-maximizing auction in which in every instance, the item is sold to at most one of the $k$ bidders with the highest values, where $k$ is ideally some small constant \cite{dobzinski-fu-kleinberg, XGPL, dobzinski-uziely, CFK, FLLT}. 
The most relevant result is that the approximation ratio of the $k$-lookahead auction approaches $\frac e {e+1}$ \cite{XGPL} and that this is tight \cite{dobzinski-uziely}. 

The literature on interim IR mechanisms is also quite rich. The stunning result here is that of Cremer and McLean \cite{CM85,CM88}, which shows that when the joint distribution satisfies a particular full rank condition, the revenue that the auctioneer can extract equals to \emph{all} the expected social welfare. This revenue can be extracted by running a simple second-price auction and charging the participating bidders appropriate fees. However, Albert, Conitzer, and Lopomo \cite{ACL} show that this result cannot be extended to all distributions by showing that there are distributions for which the ratio between the optimal revenue and the optimal welfare goes to $0$. Feldman and Lavi \cite{FL} further show that this gap exists even for distributions that have ``almost full'' rank. 

We stress that there are no known interim IR mechanisms that play an analogous role to that of the lookahead auction in the ex-post IR universe: there are no natural, simple to describe auctions that for every distribution provide a significant fraction of the optimal revenue. To put it differently, if our goal is revenue maximization, ex-post mechanism designers can offer us a rich toolbox to work with, whereas interim IR mechanism designers either offer us a dream solution (but only if our distribution happen to obey the Cremer-McLean condition), or leave us empty handed (if we want our solution to generically work for all possible distributions). A primary goal of this paper is to understand whether the set of tools of interim IR mechanism design can be significantly extended.

\subsubsection*{Understanding the Success of Cremer-McLean via Kolmogorov Complexity}

Our first set of results (Section~\ref{kc-section}) attempts to explain the singular success of Cremer-McLean in the interim IR universe. We would like to show that there are no generic, simple-to-describe allocation functions that extract a significant fraction of the optimal (interim IR) revenue for all distributions. To put some technical sense to this statement, we analyze the Kolmogorov complexity of the allocation function of good mechanisms. Recall that, informally speaking, the Kolmogorov complexity (see, e.g., \cite{li2008introduction}) of a string is the size of the smallest Turing machine that generates it. We are interested in the Kolmogorov complexity of functions, so we view an allocation function as a string over the alphabet $[n]$ that its $i$'th position specifies which of the $n$ players receives the item in the $i$'th instance. Our results hold whenever the indices of the string correspond to instances that are ordered in some ``natural'' order. Roughly speaking, an order is natural if it has the almost obvious property that there exists a small Turing machine that given $i$, prints the $i$'th instance. 

When the distribution obeys the Cremer-McLean condition, the optimal allocation function is simple: give the item to a highest-value player. It is not hard to see that this implies that the Kolmogorov complexity of the optimal mechanism is low (logarithmic in the size of the support). In contrast, for some distributions for which the Cremer-McLean condition does not hold, we prove that even if we settle on approximations, the allocation function has high Kolmogorov complexity.

\vspace{0.1in} \noindent \textbf{Theorem: } Fix some constant $0<c<1$. There are distributions for which the Kolmogorov complexity of the allocation function of every mechanism that guarantees a $c$-fraction of the revenue-maximizing interim IR mechanism is polynomial in the size of the support.

\subsubsection*{Understanding the Limitations of Interim IR Mechanism Design via Rigidity}

We prove the last theorem by introducing and studying the the novel concept of \emph{rigidity} (Section~\ref{rigid-sec}). Informally, $S$ is rigid in $\mathcal F$ if for every mechanism $M'$ with an allocation function $f'$ where $f$ and $f'$ agree on the allocation of at most $x$-fraction of the instances of $S$, it holds that the expected revenue of $M'$ is at most an $x$ fraction of the optimal revenue.\footnote{As stated here, there is a linear dependency between $x$ and the approximation ratio, but the definition and our treatment are more general.}

We study rigidity under both notions of individual rationality. First, observe that rigid sets can be easily constructed: suppose that player $1$ always has some very high value and the other bidders always have very low values. For simplicity assume a uniform distribution on the instances. The optimal mechanism always sells player $1$ the item at high price. Clearly, any mechanism that allocates that item to player $1$ in at most fraction $x$ of the instances extracts at most $x$ fraction of the optimal solution, hence the set of all instances is rigid with respect to this distribution.

Assuming ex-post individual rationality, Ronen's lookahead auction shows that in every rigid set the highest value player should be allocated the item at least half of the time, simply because only the player with the highest value can be allocated in the lookahead auction and because the lookahead auction extracts at least half of the optimal revenue. In contrast, assuming interim individual rationality, we show that there are distributions with rigid sets in which the optimal allocation function  do not have a simple pattern. For example, the revenue that the optimal mechanism extracts from players with the $i$'th highest value almost equals to the revenue that is extracted from players with the highest value, for all values of $i$. This is in fact a corollary of the previous theorem that more precisely shows that the Kolmogorov complexity of the allocation function is high, even when restricted to the rigid sets.

Since the existence of rigid sets essentially implies that the hands of the designer are tied, our results help explain why we have little hope of developing generic constructions (a-la Ronen's ex-post lookahead auction) in the interim individual rationality world.


\subsubsection*{Examples for Rigid Sets}



We now describe two distributions with complex rigid sets. Both are obtained and analyzed using our main technical result, which is a generic way of obtaining complex rigid sets (Section~\ref{construction_sec}). We give a high-level description of our approach. We start with some set of instances $S$ and construct ``unnatural'' allocations on it. We then use this set to define a probability distribution $ \mathcal F$ over a larger set of instances. The support of the distribution $ \mathcal F$ includes instances that are not all in $S$. Consider now an incentive-compatible mechanism for the distribution $ \mathcal F$. Define the \emph{agreement ratio} of the mechanism $M$ with the allocation on the set of instances $S$ to be the fraction of the instances of $S$ such that the allocation function of the mechanism $M$ coincides with the unnatural allocation function. Our key finding is that $S$ is rigid: the revenue of $M$ is at most $\alpha+\eps$ of that of the optimal mechanism for the distribution $ \mathcal F$, where $\alpha$ is the agreement ratio and $\eps>0$ is very small. 

This paper provides two constructions of complex rigid sets (Section~\ref{ex-rigid-interim}). We now give an imprecise description of each construction. The allocation function is deterministic in both constructions, but the construction process is random. In both constructions, we focus on making the player that is allocated the item indistinguishable from a large fraction of the other players.
\begin{enumerate}
\item \textbf{Random High Values} (Section~\ref{first_method_sec}). We construct some base set of values, where about half of the players have values that are distributed uniformly and independently between $\frac 1 2$ and $1$ and the other players have values very close to $0$. From this base set we generate $\frac n 2$ instances, where in each such instance, one of the remaining $\frac n 2$ players (the ``active'' player) has some value that is uniformly distributed between $\frac 1 2$ and $1$. All other values are identical to their value in the base set. The partial allocation function allocates the item in each of these $\frac n 2$ instances to the active player. We repeat this process $m$ times (for some very large $m$), each time starting with a different base set, so in total, our partial allocation function is defined on $m\cdot \frac n 2$ instances. Without taking a ``global'' view of the allocation function, it is hard to make sense of it: in every one of the instances, we expect $\frac n 2+1$ to have values that are distributed between $\frac 1 2 $ and $1$, and we allocate the item to an arbitrary one of them. 

\item \textbf{Geometrically increasing base sets} (Section~\ref{second_method_sec}). In this construction method, we start with some arbitrary base set of values $v_1,...,v_n$, e.g., for each $i$, $v_i=1$. From this base set we generate $n$ instances, where in the $i$'th instance the $i$'th player has value $r_i\cdot v_i$, for some $r_i$ that is chosen uniformly at random from $[2,4]$, say. Player $i$ is allocated the item in that instance. We repeat this process $m$ times, for some large $m$, each time the base set is obtained from the previous base set by multiplying each $v_i$ by some $r_i$ that is chosen independently and uniformly at random from $[2,4]$. Note that after not too many repetitions, the instances will look ``random'', as the value of the player that is allocated the item is essentially indistinguishable from the values of the rest of the players. All players are considered active for each base set in this construction method.
\end{enumerate}

We show that each of the partial allocation functions constructed by the above two methods can be ``embedded'' into some distribution $\mathcal F$ and that the approximation ratio of any mechanism deteriorates in essentially a linear fashion with the agreement ratio of the allocation function of the mechanism and the partial allocation function that we started with.

We note that the two constructions methods are just two concrete applications of a more general lemma (\ref{construction_thm}). This lemma takes some allocation function on a set of instances $S$, embeds the set $S$ in a distribution $ \mathcal F$, and gives an upper bound on the fraction of the optimal revenue that can be extracted in terms of the agreement ratio with $S$ and the structure of $S$. Most of the technical difficulty in this paper is in proving this general lemma.

\section{Preliminaries}

We have one seller, one indivisible item and $n$ bidders. Each bidder has a private value $v_i \in D_i$ for the item and these values are drawn from a joint distribution $F$ over the set of possible instances $D= D_1 \times \dots D_n$.

A (deterministic) \emph{mechanism} $M$ is a tuple $M=(x,p)$ where $x:D \to \{0,1\}^n$ is the allocation function and $p:D \to \mathbb{R}^n$ is the payment function. The allocation function satisfies the feasibility constraint: $\sum\limits_{i=1}^n x_i(v) \leq 1 $ for every $v \in D$. 

A mechanism $M=(x,p)$ is  \emph{dominant strategy incentive compatible} if for every player $i$, every $v_{-i} \in D_{-i}$, and every $v_i, v'_i \in D_i$ it holds that $ x_i(v_i,v_{-i}) \cdot v_i -p_i((v_i,v_{-i}) \geq  x_i(v'_i,v_{-i}) \cdot v_i -p_i((v'_i,v_{-i})$. 

The LHS of the last inequality is the \emph{profit} of bidder $i$ (w.r.t. $M$) and is denoted $\pi^M(v_i,v_{-i})$. The \emph{expected profit} of  bidder $i$ with value $v_i$ (w.r.t. $M$) is $\mathop{\mathbb{E}}_{{v}_{-i}\sim \mathcal{F}_{-i} | v_i=v_i}[\pi^M_i(v_i,v_{-i})]$.

The \emph{expected revenue} $REV(M,\mathcal{F})$ of a mechanism $M$ in the distribution $\mathcal F$ is the expected sum of the payments of the bidders.

\begin{definition}
A mechanism $M=(x,p)$ is \emph{interim individually rational} if for every $i \in [n]$ and every $v_i \in D_i$, the expected profit of bidder $i$ who knows only his own value (and the underlying distribution $F$) and bids truthfully is non-negative, i.e., $\mathop{\mathbb{E}}_{{v}_{-i}\sim D_{-i} | v_i=v_i}[\pi_i^M(v_i,v_{-i})] \geq 0$.

A mechanism $M=(x,p)$ is \emph{ex-post individually rational} if for every $i \in [n]$ and every $v \in D$, the ex-post profit of player $i$ who bids truthfully is non-negative, i.e., $\pi_i^M(v_i, v_{-i}) \geq 0$.
\end{definition}

\begin{definition}
The \emph{approximation ratio} of an interim IR, deterministic and dominant strategy incentive compatible mechanism $M$ w.r.t a distribution $\mathcal F$ is at most $\beta \leq 1$ if:
$$
\frac{REV(M,\mathcal{F})}{REV(OPT,\mathcal{F})} \leq \beta
$$
where $OPT$ is an interim IR, deterministic and dominant strategy incentive compatible mechanism that maximizes the revenue of $\mathcal F$. 
\end{definition}


\begin{definition}\label{parial-func-set}
A partial monotone allocation function is a monotone allocation function that is defined for a subset of all instances and can be extended to all instances by a monotone function. 

A partial monotone allocation function $f$ can be specified by a \emph{partial monotone allocation set}: this is a set whose elements are tuples $\{(\Vec{v}, i) \}$ such that $f(\Vec{v})=i$ and if no player is allocated the in $\Vec{v}$, we write $f(\Vec{v}) = 0$. 
\end{definition}

Throughout this document we refer to Cremer and McLean's condition for a distribution, we give  a formal definition for it, but before that we need to define the conditional probability matrix of a bidder. 

\begin{definition} \label{cond_prob_matrix}
The conditional probability matrix of bidder $i$ w.r.t a discrete distribution $\mathcal{F}$ over a domain $D$ is a matrix $CP_i(\mathcal{F})$ of dimensions $|D_i| \times |D_{-i}|$ where for every $ 1 \leq k \leq |D_i|$ and every $1 \leq j \leq |D_{-i}|$ we have: 
$$
[CP_i(\mathcal{F})]_{(k,j)} = \Pr_{v \sim F}(v_{-i}=v^j_{-i}| v_i=v_i^k)
$$ 
\end{definition}

\begin{definition} \label{full-rank-cond}
A distribution $\mathcal{F}$ over the set of of instances satisfies the \emph{full rank condition, or Cremer and McLean's condition} if for every player $i$, his conditional probability matrix $CP_i(\mathcal{F})$ has full rank.
\end{definition}

\section{Rigidity} \label{rigid-sec}

In this section we introduce the notion of Rigidity. Roughly speaking, a distribution $\mathcal{F}$ is rigid with respect to a partial monotone allocation set $S$ if the approximation ratio of every mechanism $M$ for the distribution $\mathcal{F}$ depends on the fraction of instances in which $M$ allocates the item as in $S$.



We start this section with definitions and notations for rigidity and some basic properties of it (Section~\ref{per-rigid}), then in Section~\ref{app-claims}, we describe and prove several structural results of rigid sets. 


\subsection{Definitions and Basic Properties} \label{per-rigid}



We now define the notion of rigidity and study some basic properties of it. First, we require the following definitions:

\begin{definition} \label{disagreement_ratio}
The \emph{disagreement ratio} of a mechanism $M$ with some partial allocation multi-set $S$ is the fraction of instances in $S$ for which the allocation of $M$ is different than the allocation of $S$.
\end{definition}

\begin{definition}\label{rev-dis-func}
A \emph{revenue disagreement function} is a function $f:[0,1] \to [0,1]$ that is monotone non-increasing.
\end{definition}

We are now ready to define rigidity:

\begin{definition}[Rigidity]\label{rigid-def}
Let $\mathcal F$ be a distribution over a set of instances $D$, $S$ a partial monotone allocation multi-set over $D$, and
$f$ some revenue disagreement function.
Then, $\mathcal F$ is ex-post (interim) IR $f$-rigid with respect to $S$ if the approximation ratio of every dominant strategy incentive compatible, deterministic and ex-post (interim) IR  mechanism $M$ with disagreement ratio of at least $x$ with $S$ is at most $f(x)$.   
\end{definition}

We sometimes abuse notation and say that a partial monotone allocation set $S$ is ex-post (interim) IR  rigid. By that, we mean that there is a distribution and a revenue disagreement function $f$ such that $\mathcal F$ is ex-post (interim) IR $f$-rigid with respect to $S$, where $f$ and $\mathcal F$ are clear from the context.

In the definition of rigidity, $S$ was allowed to be a multi-set so that different allocations could affect the disagreement ratio of a mechanism differently. Also, $S$ may have instances where no player is assigned the item.

Now, we present the main family of revenue disagreement functions that we analyze in the paper.

\begin{definition}\label{al-def}
Fix $\varepsilon \in (0,1)$. A revenue disagreement function is \emph{$\varepsilon$-almost linear} if for every $x \in (\varepsilon,1]$ it holds that $f(x) \leq  1-x +\varepsilon$ and for every $x \in (0,\varepsilon]$ it holds that $f(x) < 1$. 
\end{definition}

Observe an $\varepsilon$-almost linear revenue disagreement function is strictly smaller than 1 for every value in $(0,1]$. Therefore, even a disagreement of one instance implies that the revenue is smaller than the optimal revenue. 
Also observe that an $\varepsilon$-almost linear revenue disagreement function is also $\varepsilon'$-almost linear revenue disagreement function for every $ \varepsilon < \varepsilon' < 1$.

\subsubsection{Union of Rigid Sets} \label{union-sec}

We now study a basic property of rigid sets. This set of results is not used directly in the technical parts of the paper, but will be helpful in understanding the concept of rigidity.

\begin{claim} \label{union-claim}
Let $\mathcal{F}$ be a distribution over set $V$ of instances, $f_1, f_2$ revenue disagreement functions and $S_1, S_2$ partial monotone allocation sets over $V$. 
Assume $\mathcal{F}$ is ex-post (interim) IR $f_1$-rigid w.r.t $S_1$ and ex-post (interim) IR $f_2$-rigid w.r.t $S_2$. Let $x_1 :=  \max \{x \in [0,1]\,|\, f_1(x)=1\}$, $x_2 := \max \{x \in [0,1]\, | \, f_2(x)=1\}$. Define for every $x \in [0,1]$, $f(x) = \max \{f_1(x \cdot (1-x_1)), f_2(x \cdot (1-x_2))\}$.

Then, there exists a partial monotone allocation set $S$ composed only of elements from $S_1$ and $S_2$ such that $\mathcal{F}$ is $f$-rigid w.r.t $S$ and that at least $1-x_1$ and $1-x_2$ fractions of the elements from $S_1$ and $S_2$ respectively are in $S$.  
\end{claim}


Observe that the claim is tight in the sense that for some values of $S_1$ and $S_2$, there is no partial \emph{monotone} allocation set that contains more than $1-x_1$ fraction of the elements from $S_1$ and at least $1-x_2$ fraction of the elements from $S_2$, as the allocations in $S_1$ might conflict with the allocations in $S_2$. 
For example, consider the distribution $\mathcal{F}$ supported on two instances of two players: $(1,1)$ and $(\frac{1}{\varepsilon}, 1)$, where the first instance has probability $1-\varepsilon$ and the second has probability $\varepsilon$. 
Let $S_1 = \{((1,1),2),((\frac{1}{\varepsilon},1),2)\}$, $S_2=\{((1,1),1),((\frac{1}{\varepsilon},1),1)\}$, i.e., player $2$ gets the item in both instances in $S_1$ and player $1$ gets the item in both instances in $S_2$. Then, $f_1(0)=f_1(\frac{1}{2})=f_2(0)=f_2(\frac{1}{2}) = 1$, $f_1(1)=f_2(1)=\frac{1}{2}$ and $x_1=x_2 = \frac{1}{2}$.
Now, consider a set $S$ that is composed of elements from $S_1$ and $S_2$, for every element from $S_1$  in $S$ another element from $S_2$ cannot be added to $S$ and vice versa. 
However, we focus on the family of almost linear revenue disagreement functions (Definition~\ref{al-def}) for which $x_1 = x_2 = 0$, i.e., all the elements from $S_1$ and $S_2$ are in $S$.     


\begin{proof}[Proof of Claim~\ref{union-claim}]
Consider an optimal dominant strategy incentive compatible, deterministic and ex-post (interim) IR mechanism $O$ for the distribution $\mathcal{F}$. Let $S$ be the set of elements from $S_1 \cup S_2 $ that $O$ agrees with their allocation. Note that $S$ is a partial monotone allocation set as $O$ is a dominant strategy incentive compatible mechanism. 
Now, by the definition of rigidity, $O$ has disagreement ratio at most $x_1$ with $S_1$ where $x_1 = \max \{x \in [0,1]| f_1(x)=1\}$ and at most $x_2 = \max \{x \in [0,1]| f_2(x)=1\}$ with $S_2$. Then, at most $x_1$ fraction of $S_1$ elements are not in $S$ and at most $x_2$ fraction of $S_2$ elements are not in $S$. 
In the next lemma we prove that the set $S$ we defined is indeed $f$-rigid which concludes the proof. 

\begin{lemma} \label{pf-union-f}
Let $M$ be some dominant strategy incentive compatible, deterministic and IR mechanism with disagreement ratio $x \in [0,1]$ with $S$, then $M$ has disagreement ratio at least $ x \cdot (1-x_1)$ with $S_1$ or at least $x \cdot (1-x_2)$ with $S_2$ (or both).
\end{lemma}
\begin{proof}
Let $d$ be the number of allocations in $S$ that $M$ disagrees with, $d_1$, the number of allocations in $S_1$ that $M$ disagrees with and $d_2$ the number of allocations in $S_2$ that $M$ disagrees with. Then, $d \leq d_1 + d_2$, $x = \frac{d}{|S|}$ and we want to show that either $\frac{d}{|S|} \leq \frac{d_1}{|S_1|\cdot(1-x_1)}$ or $\frac{d}{|S|} \leq \frac{d_2}{|S_2|\cdot(1-x_2)}$ (or both). Assume to the contrary that: 

\begin{equation}
    \frac{d_1}{|S_1|\cdot(1-x_1)} <\frac{d}{|S|}  \iff d_1\cdot|S| < d\cdot|S_1|\cdot(1-x_1)
\leq (d_1+d_2)\cdot|S_1|\cdot(1-x_1)
\end{equation}
\begin{equation}
    \frac{d_2}{|S_2|\cdot(1-x_2)} <\frac{d}{|S|}  \iff d_2\cdot|S| < d\cdot|S_2|\cdot(1-x_2)
\leq (d_1+d_2)\cdot|S_2|\cdot(1-x_2)
\end{equation}

Then,
\begin{equation*}
    |S| < |S_1|\cdot(1-x_1) + |S_2|\cdot(1-x_2)
\end{equation*}
in contradiction to the construction of $S$ such that it contains at least $1-x_1$ fraction of the elements of $S_1$ and at least $1-x_2$ fraction of the elements of $S_2$.
\end{proof}
\end{proof}

We can further strengthen this claim when the functions are $\varepsilon$-almost linear:

\begin{corollary}\label{union-rigid-almost-linear}
Let $\mathcal F$ be a distribution that is $f$-rigid with respect to some set $S$ and an $\varepsilon$-almost linear revenue disagreement function $f$. Then, there exists a rigid set $S_c$, such that every other rigid set $S'$ of $\mathcal{F}$ with $f$, belongs to $S$ (i.e., $S' \subseteq S_c$). 
\end{corollary}


\subsection{On the Structure of Ex-post IR Rigid Sets} \label{app-claims}

In this section, we show that the family of Lookahead auctions restricts the set of ex-post IR rigid sets. We then study how Cremer and McLean's result restricts interim IR rigid sets. We show that for distributions that satisfy Cremer and McLean's condition, the only rigid sets are essentially those in which the highest player always gets the item. This section mostly considers $\varepsilon$-almost linear revenue disagreement functions, and so some of our results are stated for this class.

\begin{claim}\label{lk-rigid}
Consider a distribution $\mathcal{F}$ that is ex-post IR $f$-rigid w.r.t $S$. Let $x$ be the fraction of the allocations in $S$ to players that are not the highest player. Then, $f(x) \geq \frac{1}{2}$. 
\end{claim}

\begin{proof}
Consider the Lookahead auction \cite{ronen-lookahead}, this auction can only sell the item to the highest player and so its disagreement ratio with $S$ is at least $x$. The Lookahead auction has an approximation ratio of at least $\frac{1}{2}$ when considering dominant strategy incentive compatible and ex-post IR auctions. Therefore, $f(x) \geq \frac{1}{2}$.  
\end{proof}

\begin{corollary}\label{lk-rigid-almost-linear}
Consider a distribution $\mathcal{F}$ that is ex-post IR $f$-rigid w.r.t $S$, for an
$\varepsilon$-almost linear revenue disagreement function (for some $\varepsilon \in (0,\frac{1}{2})$). Then, the fraction of the allocations in $S$ to players that are not the highest player is at most $\frac{1}{2} + \varepsilon$.  
\end{corollary}

We can use analogous claims by considering the $2$-lookahead auction:

\begin{claim}\label{2-lk-rigid}
Consider a distribution $\mathcal{F}$ that is ex-post IR $f$-rigid w.r.t $S$. Let $x$ be the fraction of the allocations in $S$ to players that are not the highest or second highest players, then $f(x) \geq \frac{\sqrt{e}}{\sqrt{e}+1} \approx 0.622$. 

\begin{corollary}[Claim~\ref{2-lk-rigid}]\label{2-lk-rigid-almost-linear}
Consider a distribution $\mathcal{F}$ that is ex-post IR $f$-rigid w.r.t $S$, for an $\varepsilon$-almost linear revenue disagreement function (for some $\varepsilon \in (0,0.622)$). Then, the fraction of the allocations in $S$ to players that are not the highest or second highest players is at most $1-\frac{\sqrt{e}}{\sqrt{e}+1} + \varepsilon \approx 0.377 + \varepsilon$.
\end{corollary}

\end{claim}

\begin{proof}[Proof of Claim~\ref{2-lk-rigid}]
The deterministic 2-Lookahead auction is a generalization of the Lookahead auction and it can only sell the item to one of the 2 highest players. Then, its disagreement ratio with $S$ is at least $x$. The approximation ratio of the 2-Lookahead was bounded by $\frac{3}{5}$ by \cite{dobzinski-fu-kleinberg} and later improved to $\frac{\sqrt{e}}{\sqrt{e}+1} \approx 0.622$ by \cite{XGPL}. Therefore, $f(x) \geq \frac{\sqrt{e}}{\sqrt{e}+1} \approx 0.622$.
\end{proof}


We now consider interim IR mechanisms and distributions that satisfy Cremer and McLean's full rank condition.

\begin{claim}\label{CM-rigid}
Consider a distribution $\mathcal{F}$ that satisfies Cremer and McLean's full rank condition and assume there exist some revenue disagreement function $f$ and a partial monotone allocation set $S$ such that $\mathcal{F}$ is interim IR $f$-rigid w.r.t $S$. 
Let $x$ be the fraction of the allocations in $S$ that are not to the highest player, then $f(x) = 1$.
\end{claim}

\begin{proof}[Proof of Claim~\ref{CM-rigid}]
Cremer and McLean \cite{CM88} show that when the distribution satisfy some condition (full rank of the conditional probability matrix), the second price auction with some lotteries has an optimal revenue equal to the social welfare. Then, their auction has a disagreement ratio of at most $x$ with $S$ and hence $f(x) = 1$.
\end{proof}

\begin{corollary}\label{CM-rigid-almost-linear}
Consider a distribution $\mathcal{F}$ that satisfies Cremer and McLean's full rank condition. Suppose that there is some $\varepsilon$-almost linear revenue disagreement function $f$ and a partial monotone allocation set $S$ such that $\mathcal{F}$ is interim IR $f$-rigid w.r.t $S$. 
Then, all the allocations in $S$ are to the highest player.
\end{corollary}

In Claim~\ref{lk-rigid} and Claim~\ref{2-lk-rigid} we saw that ex-post IR rigid sets have a certain structure.

The next claim show that the same is not true for interim IR rigid sets. 


\begin{claim} \label{no-interim-structure}
For every $k<n$, there exists an interim IR rigid set $S_k$ with an $\frac{1}{n-k}$-almost linear revenue disagreement function, such that no allocation in $S$ is to one of the $k$-highest players. 
\end{claim}

Observe that for ex-post IR rigid set it is not possible. Every ex-post IR rigid set $S$ with an $\varepsilon$-almost linear (for some $\varepsilon < \frac{1}{2}$) revenue disagreement function has \textbf{at most} $\mathbf{\frac{1}{2} + \varepsilon}$ allocations to players that are not the highest (Corollary~\ref{lk-rigid-almost-linear}). While we show for every $\varepsilon > 0$ an interim IR rigid set $S_\varepsilon$ with an $\varepsilon$-almost linear revenue disagreement function in which all allocations are not to the highest player.

Similarly, this also shows that Corollary~\ref{2-lk-rigid-almost-linear} does not hold for interim IR rigidity. I.e., every ex-post IR rigid set $S$ with an $\varepsilon$-almost linear (for some $\varepsilon < 0.622$) revenue disagreement function has at most $\mathbf{0.377 + \varepsilon}$ allocations to players that are not the two highest players (Corollary~\ref{lk-rigid-almost-linear}). While we show for every $\varepsilon > 0$ an interim IR rigid set $S_\varepsilon$ with an $\varepsilon$-almost linear revenue disagreement function in which all allocations are not to the two highest player.

The proof of Claim~\ref{no-interim-structure} is deferred to Appendix~\ref{appendix-pf-no-interim-structure}.

\section{Kolmogorov Complexity} \label{kc-section}

In this section we analyze the Kolmogorov complexity of optimal and approximately optimal allocation functions. We start with defining some needed basic objects (Section~\ref{per-kc-sec}). Then, we show that for distributions that satisfy the Cremer and McLean's condition (Definition~\ref{full-rank-cond}) the Kolmogorov complexity of the allocation function of an optimal auction is low (Section~\ref{kc-cm-sec}). In Section~\ref{kc-rigid-general} we show the existence of distributions (that do not satisfy Cremer and McLean's condition) for which the Kolmogorov complexity of the allocation function of every mechanism with a constant approximation ratio is high.

\subsection{Natural Orders} \label{per-kc-sec}

We would like to analyze the Kolmogorov complexity of allocation functions. However, Kolmogorov complexity is defined on strings, not functions. Thus, we represent allocation functions as strings with alphabet $[n]$. Index $i$ of the string corresponds to the $i$'th instance in the support according to some full order $\prec{}$. We will consider \emph{natural} orders.
We say that an order is natural if the Kolmogorov complexity of printing all the instances in the support when they're sorted by this order is logarithmic in the size of the support. Note that this is equivalent of having a small Turing machine that gets $i$ as input and prints the corresponding instance. This seems to be a minimal requirement for ``interpreting'' the allocation function (if it is hard to understand what instances some of the positions correspond to, the representation is almost useless).

\begin{definition}\label{natural-order}
A full order $\prec$ over a domain $D$ of instances of $n$ players is \emph{natural} if the Kolmogorov complexity of the string that equals to the concatenation of all instances in $D$ sorted by $\prec{}$ is $O(\log|D|)$.
\end{definition}

For example, the lexicographic order is natural in the domain $\{1, \dots, H\}^n$. 
To see this, consider the program that starts with a string $0\dots 0$ and each time increases its value by one (as if it were a number in base $H$) and prints it.

We now define a specific order $\cmorder$. We will analyze the Kolmogorov complexity of the strings that correspond to allocations ordered by $\cmorder$ and show that it is high. Our result will then be extended to all natural orders since we will show that transforming every string ordered by a natural order to a string ordered by $\cmorder$ can be done by a small Turing machine.

\begin{definition} \label{cm-order}
Let $\cmorder$ be the following order over domain $D$ of instances of $n$ players:
instance $v \in D$ is before instance $v' \in D$ if $\argmax_{i \in [n]}\{v_i\} \leq \argmax_{i \in [n]}\{v'_i\}$. In case of equality, we break ties by lexicographic order (i.e., let $j$ be the first index in which $v$ and $v'$ differ, then $v_j \prec {v'}_j$).
\end{definition}

The next two claims show that a lower bound on the Kolmogorov complexity of strings ordered by $\cmorder{}$ implies a lower bound on the Kolmogorov complexity of strings ordered by some natural order.

\begin{claim} \label{low-kc-comperator-for-cm-order}
Fix some domain $D$ of instances of $n$ players.
The encoding of a Turing machine that gets two instances from $D$ and determines their order according to $\cmorder$ is at most $c_2\cdot n$, for some constant $c_2$.
Specifically, given two instances $v_1, v_2$, the Turing machine returns $0$ if $v_1 \cmorder v_2$ and $1$ otherwise.
\end{claim}

\begin{proof}[Proof of Claim~\ref{low-kc-comperator-for-cm-order}]
The operations of copying a string and comparing two numbers can be implemented using Turing machine with a linear number of states in the size of the input alphabet (which in our case is $n$). Consider the following Turing machine: first, the machine copies the two input strings ($v_1, v_2$). For their first copy, we find the maximal value in each string and save its index. If the index of the first input is smaller (larger) than the index of the second input then $v_1 \cmorder v_2$ ($v_1 \cmordersucc v_2$) and we print $0$ ($1$). If the two indices are equal we continue to the next stage. This stage takes at most $O(n)$ states.
In the second stage, we work on the second copy of our input strings and compare the two strings until we find the first index that they are different on, if the value of $v_1$ in this index is smaller than $v_2$'s value in this index, we print 0. Otherwise we print 1. This stage takes at most $O(n)$ states as well.
\end{proof}


\begin{claim}\label{natural-to-cm}
Fix a distribution $\mathcal{F}$ over a domain of instances $D$. Let $\prec{}$ be some natural order over the instances in $D$ and  $\cmorder$ be the order specified in~\ref{cm-order}. Let $f$ be an allocation function for the distribution $\mathcal{F}$, and $s$ and $s'$ be the allocation strings of $f$ with respect to the orders $\prec{}$ and $\cmorder$ respectively over the instances in $D$.
Suppose that the Kolmogorov complexity of $s'$ is at least $m$. Then, the Kolmogorov complexity of $s$ is at least $m -c_1\log|D| - c_2n -c_3$ for some constant $c_3$, a constant $c_1$ that depends only on the order $\prec{}$ and a constant $c_2$ that only depends on the order $\cmorder$. 
\end{claim}

\begin{proof}[Proof of Claim~\ref{natural-to-cm}]
Assume Claim~\ref{natural-to-cm} does not hold. Then, there exists a distribution $\mathcal{F}$ over a domain of instances $D$, $f$ an allocation function for the distribution $\mathcal{F}$, $s$ and $s'$ that are the allocation strings of $f$ with respect to the orders $\prec{}$ and $\cmorder$ respectively over the instances in $D$, such that the Kolmogorov complexity of $s'$ is at least $m$ but the Kolmogorov complexity of $s$ is less than $m -c_1\log|D| - c_2n -c_3$.

Let $P_1$ be a Turing machine that prints $s$ with encoding size less than  $m -c_1\log|D| - c_2n -c_3$.
Consider the following program for printing $s'$.
First, generate $s$. Second, generate all the instances in $D$ by order $\prec{}$. Then sort all the instances in $D$ according to the order $\cmorder$ and at the same time sort the string $s$ by the same swaps that are made for the instances in $D$ and print the sorted $s$. 

We now explain how to implement this program while maintaining encoding size of less than $m$, which will contradict our assumption. 
For the first step we use the program $P_1$ whose encoding size is less than $m -c_1\log|D| - c_2n -c_3$. For the second step, since $\prec{}$ is a natural order, there exists a Turing machine $P_2$ whose encoding is at most some $c_1\cdot \log |D|$ and it simulation gives us all instances in $D$ ordered by $\prec{}$. For the third step, we need two components, the first is a ``comparator'' function $P_3$ for the order $\cmorder$ that gets two instances and returns their order according to $\cmorder$ and has an encoding of size at most some $c_2 \cdots n$. The second component is a sorting algorithm with small encoding, we use, say, Bubble Sort. 
The sorting algorithm will simultaneously sort the string s by using same swaps that it made for the instances in the list. Observe that bubble sort needs a constant number of states and together with $P_3$ we get an encoding which is linear in $n$  
Now, this program has KC of at most $k + c_1\log|V| + c_2\cdot n +c_3$, where $c_3$ is some constant. This contradicts our assumption, as needed.
\end{proof}

Thus, from now on we will analyze the Kolmogorov complexity of the order $\cmorder$ and get, as a corollary, the Kolmogorov complexity of any natural order.


\subsection{Low Kolmogorov Complexity of Cremer and McLean's Distributions}\label{kc-cm-sec}

In this section we state and prove the theorem about the low Kolmogorov complexity of the allocation function of an optimal auction of distributions that satisfy the Cremer and McLean's condition (see Definition~\ref{full-rank-cond}). We consider the Kolmogorov complexity of the allocation function with respect to the order $\cmorder$ (definition \ref{cm-order}).  

\begin{theorem}\label{CM-kc}
Let $\mathcal{F}$ be a distribution that satisfies the Cremer and McLean's condition with support size $k$. There exists an optimal auction for $\mathcal F$ for which the Kolmogorov complexity of the allocation function is $O(n\log k)$ with respect to 
every natural order.
\end{theorem}

\begin{proof}[Proof of Theorem~\ref{CM-kc}]
Cremer and McLean showed that the optimal auction for every distribution that satisfies the full rank condition (Definition~\ref{full-rank-cond}) is a second price auction with appropriate fees. Recall that the allocation function of the second price auction simply gives the item to a maximum-value player.
Using the natural order we can retrieve all the instances in the support (in logarithmic Kolmogorov Complexity), then we go over them one by one and find the player with the maximal value in each instance, and this player gets the item.

\end{proof}

Observe that the Kolmogorov complexity of Myerson's optimal mechanism for independent distributions \cite{Myerson81} is also logarithmic in the size of the support under natural orders: using the natural order we can go over all the instances in the support one by one and compute in each of them the virtual values of each player. The player with the maximal virtual value gets the item.

\subsection{The Kolmogorov Complexity of Approximations} \label{kc-rigid-general}

In this section we use rigid sets to analyze the Kolmogorov complexity of the allocation function of every mechanism that provides a good approximation ratio. The notion of a rigid set allows us to focus on the allocation of a mechanism in the instances of $S$, and so, if the mechanism has a good approximation, its allocation string restricted to the instances of $S$ has to be relatively close to the allocation string of $S$.

\begin{theorem}\label{kc-app-rigid}
Consider distributions $\mathcal{F}_1, \dots, \mathcal F_k$, all with support size of at most $r$. Suppose each $\mathcal F_i$ is $f$-rigid with respect to $S_i$, where $f$ is some $\varepsilon$-almost linear revenue disagreement function (for some $\varepsilon \in [0,1)$). 
Suppose that all the following conditions hold for some $c \in (\varepsilon,1)$, $\prec{} $ a full order, and $\kcparam \in \mathbb{N}$:
\begin{enumerate}
    \item \label{same-size} $|S_1|=|S_2|=\cdots=|S_k|$, and let $g= |S_1|$.
    \item \label{diffrent-alloc} For every $S_j$, $\prec$ imposes an order over the instances in $S_j$. Let $s_j$ be the allocation string for the instances in $S_j$ when they are sorted by this order and their allocations are as specified in $S_j$. Then, for every $i \neq j$ we have $s_i \neq s_j$. 
    \item \label{good-big-bad} 
    Let $x = 1- c + \varepsilon$, then $\frac{k}{\binom{r}{g}} > n^\kcparam \cdot \binom{g}{x \cdot g}\cdot n^{x\cdot g}$.
\end{enumerate}
Then, there exists some distribution $\mathcal F_j$, such that the Kolmogorov complexity of the allocation function of every auction that provides a $c$ approximation for $\mathcal F_j$ is at least $\kcparam$ with respect to the order $\prec{}$.
\end{theorem}

\begin{proof}[Proof of Theorem~\ref{kc-app-rigid}]

For every $j \in [k]$, consider an allocation string $a$ for the distribution $\mathcal F_j$. $a$ has exactly $g$ indices that correspond to the instances from $S_j$ and $a$ is of size at most $r$ so there are at most $\binom{r}{g}$ different subsets of $g$ indices that can correspond to the instances of the set $S_j$. This is true for every $j \in k$ and so at least $\frac{k}{\binom{r}{g}}$ of the distributions $\mathcal F_1, \dots, \mathcal{F}_k$ share the exact same set of indices that correspond to the instances in their respective rigid sets $S_1, \dots S_k$, we denote the set of these distributions by $O$ and the set of the indices by $I$. Observe that by property~\ref{diffrent-alloc}, for every $\mathcal F_i, \mathcal F_j \in O$ with $i \neq j$ it holds that $s_i \neq s_j$.


\begin{lemma} \label{single-optimal}
Fix an allocation string $b$. $b$ corresponds to an allocation function of an optimal mechanism of at most one of the distributions in $O$.  
\end{lemma}

\begin{proof}[Proof of Lemma~\ref{single-optimal}]
Since $f$ is an $\varepsilon$-almost linear revenue disagreement function, for every $\mathcal{F}_j \in \mathcal{F}_1, \dots, \mathcal F_k$ any optimal allocation must agree with all the allocations in $S_J$. Thus, for an allocation string $b$ to be an optimal allocation to more than one distribution in $O$, it has to agree with at least two different allocations on the same set of indices $I$, which is a contradiction.
\end{proof}

Denote by $L_{\kcparam}$ the set of all strings with Kolmogorov Complexity less than $\kcparam$. Let $B_{\kcparam} \subseteq O$ be the set of distributions in $O$ such that a distribution $\mathcal{F}_j \in O$ is in $B_\kcparam$ if it has a mechanism with approximation ratio at least $c$ whose allocation string has Kolmogorov complexity less than $\kcparam$. Observe that the claim follows immediately by showing that $B_\kcparam$ is a strict subset of $O$: 

\begin{lemma} \label{O-minus-B-big}
$B_\kcparam$ is a strict subset of $O$.
\end{lemma}

\begin{proof}[Proof of Lemma~\ref{O-minus-B-big}]
By Lemma~\ref{single-optimal}, each string in $L_\kcparam$ is an optimal allocation to at most one of the distributions in $O$.
The size of $L_\kcparam$ is at most the number of strings with Kolmogorov complexity less than $\kcparam$, which is $\sum_{j=0}^{\kcparam-1}n^j = \frac{n^\kcparam -1}{n-1} < n^\kcparam$, since the left hand side is the number of different Turing machines over the alphabet $[n]$ with encoding size less than $\kcparam$.
Now, for every string $a \in L_\kcparam$ we bound the number of distributions $\mathcal{F}_j$ in $O$ for which $a$ restricted to the indices in $I$ (denoted $a_I$) and the allocation string of $S_j$ are different in at most $x\cdot g$ places. We bound it by at most $\binom{g}{g\cdot x} \cdot n^{g\cdot x}$;
There are $\binom{g}{g\cdot x}$ different sets of size $g\cdot x$ of locations in the string $a_I$ of size $g$. The number of possible values in each location is $n$. Overall there are at most $\binom{g}{g\cdot x} \cdot n^{g\cdot x}$ such distributions (as each distribution in $O$ has a different allocation in the indices in $I$).
Therefore, there are at most $n^\kcparam \cdot \binom{g}{g\cdot x} \cdot n^{g\cdot x}$ distributions in $B_\kcparam$ and at least $\frac{k}{\binom{r}{g}}$ distributions in $O$. Now, by Property~\ref{good-big-bad} there are more distributions in $O$ than in $B_\kcparam$. 
\end{proof}
\end{proof}

\subsubsection{Distributions with High Kolmogorov Complexity for Every Approximation} \label{kc-high-no-cm}

Now that we have the right tools prepared, we can apply a specific construction to prove that there exist distributions for which the allocation function of every approximation mechanism has high Kolmogorov complexity for every natural order $\prec{}$ (Definition~\ref{natural-order}). We will rely on the construction of Section \ref{set-of-m-divi-for-kc-sec} that is given later.

\begin{corollary}\label{kc-app-cor}
For every constant $0 < c < 1$ and every large enough $n,m \in \mathbb{N}$, there exists a distribution with support linear in $m\log^2 n$, for which the Kolmogorov complexity of the allocation function of every mechanism with approximation ratio $c$ is at least $m$ for every natural order.
\end{corollary}

\begin{proof}[Proof of Corollary~\ref{kc-app-cor}]
We use the construction of Section \ref{set-of-m-divi-for-kc-sec}. We show that for every fixed $0 < \varepsilon < c$ and large enough $n,m$, Theorem~\ref{kc-app-rigid} holds with $ S_1, \dots, S_k $ as the sets in $R_{n,m}$ (defined in Section~\ref{set-of-m-divi-for-kc-sec}), $\mathcal F_1, \dots,  \mathcal{F}_k$ as the distributions guaranteed by Theorem~\ref{construction_thm} when given $m$-divisible set from $R_{n,m}$, $f$ as their $\varepsilon$-almost linear revenue disagreement function (from Corollary~\ref{rigid_construction}), $\cmorder$ as the order $\prec{}$, and $\kcparam=l\cdot m$ for some constant $l$ to be specified later. 

First, by Claim~\ref{properties-of-R-n-m}, each $m$-divisible set $S$ in $R_{n,m}$ is of size $m \cdot \log n$, satisfies the uniqueness of thresholds property (~\ref{uniqueness}) and each of its $m$ subsets has a set of $\log n$ active players.
Then $g= m\log n$ and by Theorem~\ref{construction_thm} $r \leq 6m\log n + m\log^2 n$. Hence, Property~\ref{same-size} holds. 
By Claim~\ref{properties-of-R-n-m}, the parameter of an $m$-divisible set $S \in R_{n,m}$, $c_S$ is at most $c_S \leq \frac{4n}{m\cdot \log n} + \frac{4}{\log n}$ and thus by Corollary~\ref{rigid_construction} $f$ is an $\varepsilon$-almost linear revenue disagreement function when $\frac{4n}{m\cdot \log n} + \frac{4}{\log n} \leq \varepsilon <c$ (which is true for large enough $n,m$).

In Lemma~\ref{ordering-r-n-m} and Lemma~\ref{alg-r_n-m} we show that Property~\ref{diffrent-alloc} and Property~\ref{good-big-bad} hold respectively.

\begin{lemma} \label{ordering-r-n-m}
For every $S_j$, $\cmorder$ imposes an order over the instances in $S_j$. Let $s_j$ be the allocation string for the instances in $S_j$ when they are sorted by this order and their allocations are as specified in $S_j$. Then, for every $i \neq j$ we have $s_i \neq s_j$. 
\end{lemma}

\begin{proof}[Proof of Lemma~\ref{ordering-r-n-m}]

Recall that $S_j$ is an $m$-divisible set with respect to its subsets $S^1_j, \dots S^m_j$ (Definition~\ref{m-divisible}). By Claim~\ref{properties-of-R-n-m}, in every instance in $S_j$, the first player has the maximal value and his value is the same for all the instances in the same subset $S^l_j$. Moreover, the values of the first player in the different $m$ subsets are strictly increasing. Hence, in the support of the distribution, when considering the order $\cmorder$ the instances that belongs to the first subset $S^1_j$ appears first and then the instances from the second subset $S^2_j$ and so on until the instance from $S^m_j$. 
Furthermore, every two instances $v,v'$ in the same subset $S^j_j$ differ from each other in exactly two places, these places correspond to the players that get the item in these instances. When player $i$ gets the item in $v$ and player $i'$ gets the item in $v'$ we have $v_i > v'_i$ and $v_{i'} < v'_{i'}$ and so the order inside each subset is a decreasing order over the indices of the players that get the item (i.e., if $i$ gets the item in $v \in S^l_j$ and $i'>i$ gets the item in $v' \in S^l_j$ then $v' \cmorder \, \, v$), the players in the active set of $S^l_j$.
Lastly,
$R_{n,m}$ (as described in Section~\ref{set-of-m-divi-for-kc-sec}) has $k =(\binom{n}{\log n})^m$ different allocations, in every two different $m$-divisible sets $S_j, S_{j'} \in R_{n,m}$ the set of active players in some repetition $j \in [m]$ is different and so $s_j \neq s_{j'}$ and Property~\ref{diffrent-alloc} holds.

\end{proof}



\begin{lemma}\label{alg-r_n-m}
Property~\ref{good-big-bad} holds for large enough $n,m$.
\end{lemma}

\begin{proof}[Proof of Lemma~\ref{alg-r_n-m}]
We need to show that for large enough values of $n,m$ it holds that:
\begin{equation*}
    \frac{k}{\binom{r}{g}} > n^\kcparam \cdot \binom{g}{xg}\cdot n^{xg}
\end{equation*}
for $k = \binom{n}{\log n}^m, g= m\log n, \, \kcparam = l\cdot m$ and $r< 6m\log n + m\log^2 n$.

We use the known bounds on the binomial coefficient:
\begin{equation*}
    (\frac{n}{k})^k \leq \binom{n}{k} \leq (\frac{n\cdot e}{k})^k
\end{equation*}

and get:
\begin{subequations} \label{new-bounds}
\begin{equation*}
    \frac{k}{\binom{r}{g}} = \frac{\binom{n}{\log n}^m}{\binom{(m\log n)(6 + \log n)}{m\log n}} \geq (\frac{n}{e\log n (6 + \log n)})^{m\log n}.\\
\end{equation*}
\begin{equation*}
          n^{l \cdot m} \cdot \binom{g}{xg}\cdot n^{xg} \leq n^{l \cdot m} \cdot (\frac{e}{x})^{xm\log n} \cdot n^{xm\log n}
\end{equation*}
\end{subequations}

Therefore it is enough to show that for large enough values of $n,m$ it holds that:
\begin{equation*}
    \frac{n}{e\log n(6+\log n)} > n^{\frac{l}{\log n}}\cdot (\frac{e}{x})^x \cdot n^x
\end{equation*}

Recall that $x< 1$ and so:
\begin{equation*}
    e^{l+1} \cdot n^x \geq n^{\frac{l}{\log n}}\cdot (\frac{e}{x})^x \cdot n^x.
\end{equation*}

Finally, for large enough values of $n$ we have (as $l$ is a constant independent of $n,m$): 
\begin{equation*}
    n^{1-x} > e^{l+2} \log n(6 +\log n).
\end{equation*}
\end{proof}

Now, we have that there is a distribution $\mathcal{F}_j \in \{\mathcal{F}_1, \dots \mathcal{F_k}\}$ for which the Kolmogorov complexity of every allocation function of every auction with approximation  ratio $c$ is at least $l \cdot m$, with respect to the order $\cmorder$.
By Claim~\ref{natural-to-cm} we can get that the Kolmogorov complexity of every allocation function of every auction with approximation ratio $c$ for $\mathcal{F}_j$ is at least $m$, with respect to every natural order. We choose $l$ to be large enough so $l\cdot m -c_1\cdot \log r -c_2\cdot n -c_3 \geq m$, where $c_1, c_2, c_3$ are constants from Claim~\ref{natural-to-cm} that depend on the specific natural order and a property of $\cmorder$, so we choose $l$ large enough to satisfy this for every natural order and the claim follows.
\end{proof}

\section{Explicit Constructions of Interim IR Rigid Sets} \label{ex-rigid-interim}

In this section we present two explicit constructions of interim IR rigid sets.
We construct interim IR rigid sets by using Theorem~\ref{construction_thm}. First, we describe a way to generate partial monotone allocation sets $S$ that have a certain structure, they are $m$-divisible sets (Definition~\ref{m-divisible}). Then, Theorem~\ref{construction_thm} provides us with a distribution $\mathcal{F}_S$ that is $f_S$ interim IR rigid with respect to $S$. The revenue disagreement function $f_S$, is a $c_S$-almost linear revenue disagreement function where $c_S$ is a parameter that depends on properties of the set $S$. 

In both constructions we can choose $c_S >0$ to be as small as we want by taking large enough values of $n$ (the number of players) and the size of the rigid set $S$.  

As mentioned, the partial monotone allocation sets that we construct have a certain structure which we call an $m$-divisible set. This term is quite technical and so we try to write the constructions in a way that is clear without the need to fully understand what an $m$-divisible set is. For that purpose we explain some notation that is used in the construction methods. 


The partial monotone allocation sets $S$ are constructed in some $m\in \mathbb{N}$ steps, each step $j \in [m]$ constructs a subset $S_j$ of $S$ such that $S = \bigcup\limits_{j \in [m]}{S_j}$. For each subset $S_j$, we have a set of 'active players', denoted $A_j$. This is the set of players that are allocated the item in some instance in $S_j$. We show that these are monotone allocation sets by proving that each constructed set $S$ is $m$-divisible, as $m$-divisible sets are monotone (see Section~\ref{prep_for_cons_sec}). 

The two methods for constructing partial allocation sets that are $m$-divisible are described next (Section~\ref{first_method_sec} and Section~\ref{second_method_sec}). We also describe a construction for a set of partial allocation sets (Section~\ref{set-of-m-divi-for-kc-sec}), this construction is based on the first construction method that constructs a single partial allocation set. This construction of sets of partial allocation sets is used to prove Corollary~\ref{kc-app-cor}.

\subsection{Construction I: Random High Values}\label{first_method_sec}

All instances in this construction consist of $\frac n 2+1$ players with values that are uniformly and independently distributed between $\frac 1 2$ and $1$. The remaining players have low values that are close to $0$. The partial allocation function allocates the item to one of the (approximately) $\frac n 2$ players with values in $[\frac{1}{2}, 1]$. Thus, by considering each instance by itself it is hard to guess which player the item should go to -- a more ``global'' view of the allocation function is necessary. Formally:

We start with some arbitrary vector $\vec{v_0}$ in which the values of all $n$ players are in $(0,1]$. 
\begin{enumerate}
    \item Each player $i$ is added to the set of active players in the $j$'th subset $A_j$ with probability $\frac 1 2$, independently at random. If the set $A_j$ is empty (which happens with probability  $\frac{1}{2}^n$), we resample it.
    \item For every active player $i \in A_j$,  sample $\Vec{r_i} \sim U[2, 4]$, and let $k_i$ be the index of the latest subset $k_i<j$ in which player $i$ was active, if no such index exists, let $k_i=0$. Then, player's $i$ value in $\vec{v_j}$ is set to:
    $$
    (\vec{v_j})_i = \frac{(\vec{v_{k_i}})_i}{r_i}. 
    $$
    \item For every non active player $i \in [n] \setminus A_j$, sample a value $v_i \sim U[\frac{1}{2}, 1]$ and set player's $i$ value in $\vec{v_j}$ to be equal to the samples value, i.e., $(\vec{v_j})_i = v_i$.
\end{enumerate}

We construct the allocations subset $S_j$ by sampling a threshold (for receiving the item when $\vec{v}_{-i} = \vec{v_j}_{-i}$) for every active player $i \in A_j$ in the $j$'th subset:
\begin{enumerate}[resume]
    \item For every player $i \in A_j$ we sample a threshold $u_{i} \sim [\frac{1}{2}, 1]$.
    \item For every player $i \in A_j$, we set the allocation in $(u_i + \thresholdepsilon, (\vec{v_j})_{-i})$ to be to player $i$ by adding the tuple $((u_i +\thresholdepsilon, (\vec{v_j})_{-i}), i)$ to $S_j$, for some arbitrary small $\thresholdepsilon > 0$.\footnote{Note that the role of $\thresholdepsilon$ is to break ties -- otherwise it is not clear who gets the item at the threshold value.}
\end{enumerate}


\begin{claim}\label{first_sampling_param}
For every $n,m \in \mathbb{N}$, the construction outputs a set $S$ of size at least $\frac{n\cdot m}{4}$ that is $m$-divisible  w.r.t. $S_1, \dots, S_m$ with
with probability at least $1- e^{(\frac{-n\cdot m}{16})}$.    
\end{claim}

\begin{observation}\label{first_method_sparse}
For every $n,m \in \mathbb{N}$, the construction method described in \ref{first_method_sec} constructs a sparse set of base vectors.
\end{observation}

\begin{proof}[Proof of Observation~\ref{first_method_sparse}]
Consider some player $i$. We will show that for every iteration $j$ player $i$'s value in the $j$'th base vector is different than player $j$'s value in every other base vector.
Suppose that player $i$ is active in the $j$'th iteration, i.e., $i \in A_j$. Then, his value in the $j$'th base vector $ \vec{v_j}_i$ is strictly smaller than his value in every other base vector $k<j$ that he is active in, and larger than every other base vector $r>j$ that he is active in. Since player $i$ is active in $j$ then $ \vec{v_j}_i < \frac{1}{2}$ and thus it is strictly smaller than his value in every other base vector $t$ that he is not active in.  

The other case is that player $i$ is not active in the $j$'th iteration. In this case we only need to argue that player $i$'s value in the $j$'th base vector is different than his value in every other base vector that he is not active in. Observe that the value of a non active player in a base vector is sampled from $U[\frac{1}{2},1]$ and thus with probability $0$ we will sample the same value more than once (as long as the number of samples is finite).  
\end{proof}

\begin{proof}[Proof of Claim ~\ref{first_sampling_param}]
The sparsity requirement (Definition \ref{sparsity}) is satisfied by Observation~\ref{first_method_sparse}. In addition every player has exactly one value $u_{i,j}$ s.t $(u_{i,j}, (v_{j})_{-i}) \in S_j$ and thus it is indeed a base set (Definition \ref{base_set}).
Therefore, $S$ is indeed an $m$-divisible set  w.r.t. $S_1, \dots S_{m}$. We now show that its parameters
are as in the statement. Define for every player $i \in [n]$ and every iteration $j \in [m]$:
$$
\mathbbm{x}_{i,j} = \begin{cases*}
        1 & \text{if $ i \in A_j $};\\
        0 & \text{otherwise.}
        \end{cases*}
$$
Then, $\mathbbm{x}_{i,j}$ is a Bernoulli random variable with expectation of $\frac{1}{2}$. Observe that the size of $S$ equals to the sum of these random variables, i.e.,  $|S| = \sum\limits_{j \in [m]} \sum\limits_{i \in [n]}\mathbbm{x}_{i,j}$. 
Using Chernoff's bound we have that the size of $S$ is less than $\frac{n \cdot m}{4}$ with probability at most $e^{(\frac{-n\cdot m}{16})}$.
Obviously, if $|S| \geq \frac{n \cdot m}{4}$, then clearly $S$ is not empty.

Recall that $g_{\text{avg}}$ is the average of $g_i = \frac{1}{1-\max\limits_{1 \leq j\leq k_i -1}\{\frac{y_{i,j}}{y_{i,{j+1}}}\}}$ over all active players and $\alpha_{\text{avg}}$ is the average of $\alpha_j= \frac{\max\limits_{i \in [n]}\{v_{i,j}\}}{\min\limits_{k \in A_j}\{u_{k,j} -v_{k,j}\}}$ over all sets. 
Therefore, $g_i = \frac{1}{1-\max\limits_{1 \leq j\leq k_i -1}\{\frac{y_{i,j}}{y_{i,{j+1}}}\}} \leq \frac{1}{1- \frac{1}{2}}=2$ and $\alpha_j= \frac{\max\limits_{i \in [n]}\{v_{i,j}\}}{\min\limits_{k \in A_j}\{u_{k,j} -v_{k,j}\}} \leq \frac{1}{\frac{1}{4}}=4$. Thus,  $\, g_{\text{avg}},\, \alpha_{\text{avg}} \leq 4$ and: $$\cfrac{ \sizeActive \cdot g_{\text{avg}}+ m\cdot \alpha_{\text{avg}}}{|S|} \leq \cfrac{4(n+m)}{\frac{n\cdot m}{4}} = \frac{16}{m}+ \frac{16}{n}.$$ 
\end{proof}

\begin{corollary}\label{agreeable_first_method}
For every small enough $\agreeableepsilon > 0$ there exists $n_\agreeableepsilon \in \mathbb{N}$ 
such that for every $n \geq n_\agreeableepsilon$ there exists 
$m \in \mathbb{N}$ for which the construction method above constructs an $m$-divisible set with $c_S \leq \varepsilon$ with high probability.

\end{corollary}

\begin{proof}[Proof of Corollary~\ref{agreeable_first_method}]
By Claim~\ref{first_sampling_param} we have that for every $n, m \in \mathbb{N}$ with probability at least $1- e^{(\frac{-n\cdot m}{16})}$ the construction method above constructs an $m' \leq m$-divisible set $S$ for which:
$$
c_S = \cfrac{ \sizeActive \cdot g_{\text{avg}}+ m\cdot \alpha_{\text{avg}}}{|S|} \leq \cfrac{4(n+m)}{\frac{n\cdot m}{4}} = \frac{16}{m}+ \frac{16}{n}.
$$

We fix a relation between $m$ and $n$, denoted by $m(n)$, so that the expression $\frac{16}{m}+ \frac{16}{n} = \frac{16}{m(n)}+ \frac{16}{n} $ approaches $0$ and $1-e^{(\frac{-n\cdot m(n)}{16})}$ approaches $1$ when $n$ approaches $\infty$. Then, for every small enough $\agreeableepsilon > 0$, there is a value $n_\agreeableepsilon \in \mathbb{N}$ such that $c_S \leq \frac{16}{m(n_\agreeableepsilon)} + \frac{16}{n_\agreeableepsilon} \leq \agreeableepsilon$. 
Hence, for every $n \geq n_\agreeableepsilon, m(n)$, we get an $m'\leq m$-divisible set with parameter $c_S \leq \agreeableepsilon$ with high probability.
\end{proof}

\subsubsection{Construction of a Set of m-divisible Sets} \label{set-of-m-divi-for-kc-sec}
In this section we describe a construction for a set of m-divisible Sets.
This construction is used in Section~\ref{kc-section} to prove Corollary~\ref{kc-app-cor}. However, this construction is very similar to the Random High Values construction and uses some of its claims, for that reason we describe this construction here.

For every large enough $n,m$, we define the set $R_{n,m}$ of $m$-divisible sets. 
Each $m$-divisible set $S=S_1, \dots S_m$ in $R_{n,m}$ will have a set of active players of size $\log n$ for every $j \in [m]$. Now, we define the set of allocations $O_{n,m}$ for m-divisible set that we want to consider and we'll make sure that for every allocation in $O$ we have an m-divisible set in $R_{n,m}$ with this allocation. 

$O_{n,m}$ is a set of size $(\binom{n}{\log n})^m$ of all possible variations of $m$ active players sets, each set of size $\log n$ and players can be from $\{2, \dots, n\}$ .


For every allocation $o \in O_{n,m}$, we construct an $m$-divisible set $S_o$ that will be added to $R_{n,m}$ in the following way:

We start with a vector $\vec{v_0} = {1}^n$.
For every $j \in [m]$, we construct the $j$'th subset $S_j$ of $S_o$ based on $S_{j-1}$ and $o_j$.
\begin{enumerate}
    \item Let $A_j = \{i \, |\, i \text{ appears in $o_j$}\}$.
    \item For every active player $i \in A_j$,  sample $\Vec{r_i} \sim U[2, 4]$, and let $k_i$ be the index of the latest subset $k_i<j$ in which player $i$ was active, if no such index exists, let $k_i=0$. Then, player's $i$ value in $\vec{v_j}$ is set to:
    $$
    (\vec{v_j})_i = \frac{(\vec{v_{k_i}})_i}{r_i}. 
    $$
    \item For every non active player $i \in [n] \setminus \{A_j \cup \{1\}\}$, sample a value $v_i \sim U[\frac{1}{2}, 1)$ and set player's $i$ value in $\vec{v_j}$ to be equal to the samples value, i.e., $(\vec{v_j})_i = v_i$.

\end{enumerate}

We construct the allocations subset $S_j$ by sampling a threshold (for receiving the item when $\vec{v}_{-i} = \vec{v_j}_{-i}$) for every active player $i \in A_j$ in the $j$'th subset:
\begin{enumerate}[resume]
    \item For every player $i \in A_j$ we sample a threshold $u_{i} \sim [\frac{1}{2}, 1)$.
    \item Sample player's $1$ value $v_1 \sim U[\max_{i \in A_j}{u_i}, 1]$
    in $\vec{v_j}$ to be the highest value among all threshold in the $j$-th iteration.
    \item For every player $i \in A_j$, we set the allocation in $(u_i + \thresholdepsilon, (\vec{v_j})_{-i})$ to be to player $i$ by adding the tuple $((u_i, (\vec{v_j})_{-i}), i)$ to $S_j$, for some arbitrary small $\thresholdepsilon > 0$.\footnote{Note that the role of $\thresholdepsilon$ is to break ties -- otherwise it is not clear who gets the item at the threshold value.}
\end{enumerate}

If the object we constructed is not an $m$-divisible set or if it doesn't satisfy the uniqueness of thresholds assumption~\ref{uniqueness} we repeat the process of construction $S_0$ (the process will stop, see Claim~\ref{properties-of-R-n-m}). Otherwise, we add $S_o$ to $R_{n,m}$. 

\begin{claim} \label{properties-of-R-n-m}
The process of constructing the set $R_{n,m}$ will stop and the following properties will hold:
\begin{enumerate}
    \item \label{uniq-holds}  Each $S \in R_{n,m}$ satisfy the uniqueness of thresholds assumption~\ref{uniqueness}. 
    \item \label{maximal-is-1} For every $S \in R_{n,m}$, in all its instances the maximal player is player 1.
    \item \label{size-of-active-set} For every $S \in R_{n,m}$, for every $j \in [m]$ the size of the set of active players in iteration $j$ is exactly $\log n$.
    \item \label{every-alloc-o} For every allocation $o \in O_{n,m}$, there exists an $m$-divisible set $S_o \in R_{n,m}$ with the allocation $o$ on its instances. 
    \item \label{param-m-divisible} Each $m$-divisible set $S$ in $R_{n,m}$ has parameter $c_S < \frac{4n}{m\log n} + \frac{4}{\log n}$
\end{enumerate}
\end{claim}

\begin{proof}[Proof of Claim~\ref{properties-of-R-n-m}]
First, we need to prove that the process halts. Assume that the process cannot generate an $m$-divisible set for some allocation $o \in O_{n,m}$ or it violates the uniqueness of thresholds assumption~\ref{uniqueness}. In this case the constructed set $S_o$ does not satisfy the sparsity condition or the uniqueness of thresholds assumption.

\begin{observation} \label{R_n-m_sparse}
For every $n,m \in \mathbb{N}$, for every $o \in O_{n,m}$ the set $S_o$ constructed by the process satisfy the sparsity condition.
\end{observation}

\begin{proof} [proof of Observation~\ref{R_n-m_sparse}]
 The proof of Observation~\ref{first_method_sparse} holds for our case as well, except for player 1. We need to show that the values of player 1 in the different base sets is different. With probability 1, the value of the maximal threshold  among the thresholds sampled in step (4) throughout the $m$ iterations is less than 1. Now, we sample the value of player 1 in every iteration $j$ from a uniform distribution over the interval $[\max_{i \in A_j}{u_i}, 1]$  and so with probability 0 we get the same value more than once. 
\end{proof}

Observe that for the uniqueness of thresholds assumption to not hold we need that the process will sample the same value twice, which happens with very small probability. 
 
Now, we show that properties $\ref{uniq-holds}-\ref{param-m-divisible}$ hold. Properties $\ref{uniq-holds}-\ref{every-alloc-o}$ hold immediately from the definition of the process. 
We analyze the parameter $c_S$ of some m-divisible set $S$ in $R_{n,m}$.
Observe that the set of all active players is of size at most $n-1$ and that $S = m\cdot \log n$. As in Claim~\ref{first_sampling_param}, $\, g_{\text{avg}},\, \alpha_{\text{avg}} \leq 4$. We get $c_S = \cfrac{ \sizeActive \cdot g_{\text{avg}}+ m\cdot \alpha_{\text{avg}}}{|S|} \leq \frac{4n +4m}{m \cdot \log n} = \frac{4n}{m\log n} + \frac{4}{\log n}$. 
\end{proof}

\subsection{Construction II: Geometrically Increasing Base Sets}\label{second_method_sec}

We start with the base vector in which all values equal $1$ and generate $n$ instances, where in the $i$'th instance the $i$'th player has value $r_i\cdot v_i$, for some $r_i$ that is chosen uniformly at random from $[2,4]$. Player $i$ is allocated the item in the $i$'th instance. We repeat this process $m$ times, for some large $m$, each time the base set is obtained from the previous base set by multiplying each $v_i$ by some $r_i$ that is chosen independently and uniformly at random from $[2,4]$. Note that after not too many iterations, the instances will look ``random'', as the value of the player that is allocated the item is essentially indistinguishable from the values of the rest of the players. All players are active in each base set in this construction method. We construct the $j$'th base vector after the previous $j-1$ base vectors:
\begin{enumerate}
    \item For every player $i \in [n]$, sample $\Vec{r_i} \sim U[2, 4]$.
    \item Define  $\Vec{v_{j}}$ by $(\Vec{v_{j}})_i = (\Vec{v_{j-1}})_i \cdot r_i$ for every player $i \in [n]$.
\end{enumerate}

We construct the allocations subset $S_j$ by sampling for every player $i$, a threshold for $\vec{v}_{-i} = \vec{v_j}_{-i}$: 
\begin{enumerate}[resume]
    \item For every player $i$ we sample a threshold $u_{i} \in [(\vec{v_j})_i \cdot 2 , (\vec{v_j})_i \cdot 4]$.
    \item For every player $i$, we set the allocation in $(u_i + \thresholdepsilon, (\vec{v_j})_{-i})$ to be to player $i$ by adding the tuple $((u_i +\thresholdepsilon, (\vec{v_j})_{-i}), i)$ to $S_j$, for some arbitrary small $\thresholdepsilon > 0$. (The addition of $\thresholdepsilon$ is for tie breaking, similarly to the previous construction.)
\end{enumerate}

\begin{claim}\label{second_sampling_param}
For every $n,m \in \mathbb{N}$, the construction method outputs a set $S = \bigcup\limits_{j \in [m]}{S_j} $ of size $n\cdot m$ that is $m$-divisible  w.r.t. $S_1, \dots, S_m$ with with parameter $c_S \leq \frac{2n + 2^m}{n \cdot m}$
\end{claim}

\begin{proof}[Proof of Claim ~\ref{second_sampling_param}]
The construction method generates m sets $S_1, \dots S_m$ such that each base vector is strictly larger than the previous ones (coordinate-wise) and thus the sparsity requirement is satisfied (Definition \ref{sparsity}). In addition every player has exactly one value $u_{i,j}$ s.t $(u_{i,j}, (v_{j})_{-i}) \in S_j$ and thus it is indeed a base set (Definition \ref{base_set}). Therefore $S$ is an $m$-divisible set  w.r.t. $S_1, \dots S_m$ and we only need to show that its parameter is as in the statement. By definition, in every set $S_j$ every player is active, hence $\sizeActive=n$ and $|S|=n\cdot m$.

Recall that $g_{\text{avg}}$ is the average of $g_i = \frac{1}{1-\max\limits_{1 \leq j\leq k_i -1}\{\frac{y_{i,j}}{y_{i,{j+1}}}\}}$ over all active players. According to this construction method $g_i = \frac{1}{1-\max\limits_{1 \leq j\leq k_i -1}\{\frac{y_{i,j}}{y_{i,{j+1}}}\}} \leq \frac{1}{1- \frac{1}{2}}=2$.

Recall that $\alpha_{\text{avg}}$ is the average of $\alpha_j= \frac{\max\limits_{i \in [n]}\{v_{i,j}\}}{\min\limits_{k \in A_j}\{u_{k,j} -v_{k,j}\}}$ over all sets. In order to analyze $\alpha_{\text{avg}}$ we analyze the sum of $\sum\limits_{j \in [m]}\alpha_j$.
Consider the $j$'th base vector $\vec{v_j}$, the maximal value of a player in it is $4^{j-1}$ and the smallest is $2^{j-1}$, and thus $\alpha_j \leq \frac{4^{j-1}}{2^{j-1}} = 2^{j-1}$. Then,  $\sum\limits_{j \in [m]}\alpha_j \leq  \sum\limits_{j \in [m]} 2^{j-1} = 2^m$, and:
$$
\cfrac{ \sizeActive \cdot g_{\text{avg}}+ m\cdot \alpha_{\text{avg}}}{|S|} \leq \frac{2n+ 2^m}{n \cdot m}.
$$
  
\end{proof}

\begin{corollary} \label{second_restricting}
For every small enough $\agreeableepsilon > 0$ there exists $m_\agreeableepsilon, n \in \mathbb{N}$ 
such that for every $m \geq m_\agreeableepsilon$ there exist $n \in \mathbb{N}$ for which the construction method above constructs an $m$-divisible set with parameter $c_S \leq \agreeableepsilon$.
\end{corollary}

\begin{proof}[Proof of Corollary ~\ref{second_restricting}]
By Claim~\ref{second_sampling_param} we have that for every $n, m \in \mathbb{N}$, the construction method above constructs an $m$-divisible set $S$ for which:
$$
\cfrac{ \sizeActive \cdot g_{\text{avg}}+ m\cdot \alpha_{\text{avg}}}{|S|} \leq \cfrac{2\cdot n+2^m}{n\cdot m}.
$$
We fix the relation between $m$ and $n$, for example, $n=2^m$, so that the expression $\frac{2\cdot n+2^m}{n\cdot m} = \frac{3}{m}$ approaches $0$ when $m$ approaches $\infty$. 
Then, for every small enough $\agreeableepsilon > 0$, there exists a value $m_\agreeableepsilon \in \mathbb{N}$ such that $\frac{3}{m_\agreeableepsilon} \leq \agreeableepsilon$. Hence, for every $m \geq m_\agreeableepsilon, n = 2^m$, we get an  $m$-divisible set with parameter $c_S \leq \agreeableepsilon$. 
\end{proof}

\section{A General Construction of Interim IR Rigid Sets} \label{prep_for_cons_sec}

In this section, we show how to ``embed'' a partial allocation function $S$ into a distribution $\mathcal{F}_S$ so that the revenue that can be extracted by a mechanism mostly depends on its agreement ratio and the structure of the allocation function, i,e., we show how to embed a set $S$ into a distribution $F_S$ so that the distribution $\mathcal{F}_S$ will be rigid with respect to $S$.
In Section \ref{ex-rigid-interim} we present some concrete partial allocation functions for which the revenue that can be extracted by incentive compatible mechanisms depends on the agreement ratio in a way that is almost linear (i.e., the revenue disagreement function is $\varepsilon$-almost linear for every small enough $\varepsilon$). We begin with some definitions.


\begin{definition}\label{parial_func}
A partial monotone allocation function is a monotone allocation function that is defined for a subset of all instances and can be extended to all instances by a monotone function. 

A partial monotone allocation function $f$ can be specified by a \emph{partial monotone allocation set}: this is a set whose elements are tuples $\{(\Vec{v}, i) \}$ such that $f(\Vec{v})=i$. 
\end{definition}

The construction of a partial monotone allocation function $f$ can be described by repeating the following process several times: choose a vector $\vec{v_j}$ that specifies the values of the $n$ players. Then, for every player $i$, choose a threshold $u_{i,j}$ (it is possible that $u_{i,j}=\infty$)
such that $f$ gives player $i$ the item in the instance $(t, (\Vec{v_j})_{-i})$ for every $t\geq u_{i,j}$.
Each vector $\vec{v_j}$ that is chosen in some iteration $j$ of the process is called a base vector.  

Observe that as described so far, this construction might not yield a feasible allocation function.
Thus, we make two changes. The first change is that in the $j$'th iteration, we require each threshold $u_{i,j}$ to be larger than $\Vec{v_j}_i$. Otherwise, if there were two players $i,i'$ with $u_{i,j}<\Vec{v_j}_i$ and $u_{i',j}<\Vec{v_j}_{i'}$, then the allocation set is not feasible since both $i$ and $i'$ have to be allocated the item in the instance $\vec{v_j}$. 

The second change is that for every two different base vectors $\Vec{v_j}, \Vec{v_k}$ and for every player $i$, we require that $(\Vec{v_j})_i \neq (\Vec{v_k})_i$. This ensures that the resulting monotone allocation function is feasible, i.e., that at most one player gets the item in every instance.
To see this, consider an instance $\vec{v}$ and suppose that the value of more than one player is higher than its threshold in this instance (without loss of generality, players $1$ and $2$). Then, for some iteration $j$ of the construction, we have that $\vec{v_j}_{-1} = \vec{v}_{-1}$ together with $u_{1,j} < \vec{v}_1$. Similarly, for some iteration $k$ of the construction, we have $\vec{v_k}_{-2} = \vec{v}_{-2}$ together with $u_{k,2} < \vec{v}_2$. Now, because whenever $j \neq k$ we required that $(\Vec{v_j})_i \neq (\Vec{v_k})_i$ for every player $i$, and in particular for player $3$ (this is why we assume $n>2$), and we have $\vec{v_j}_3 = \vec{v_k}_3$,  it must be that $j=k$. But, this means that in the same iteration, in the same base vector, we allocated the item to more than one player, which is not possible.

Based on this construction, we state some definitions. 

\begin{definition} \label{base_set}
A \emph{base set} is a partial monotone allocation set $S$ such that there exists an instance $\Vec{v_S}$ for which every $(\Vec{u}, i) \in S$ satisfies ${v_S}_{-i}= u_{-i}$. The instance $v_S$ is called the \emph{base vector} of the set. In a base set each player $i$ has at most one instance $\Vec{u}$ such that $(\Vec{u}, i) \in S$.
In addition, we require \emph{monotonicity}: for every $(\vec{u}, i) \in S$ it holds that $\vec{u}_i > \vec{v_S}_i$. 
\end{definition}

Consider some iteration $j$ of the process and observe that it defines a base set with base vector $v_j$. 



\begin{definition}\label{sparsity}
A set of base vectors $V$ is \emph{sparse} if for every two base vectors $\Vec{w}, \Vec{u}\in S, \Vec{w}\neq \Vec{u}$,  and $i \in [n]$ it holds that $w_i \neq u_i$. 
\end{definition}

By the second change, the set of all the base vectors constructed by this process is sparse. 

\begin{definition}\label{m-divisible}
A partial allocation set $S$ is \emph{$m$-divisible} with respect to $S_1, \dots, S_m$ if $S_1, \dots, S_m$ is a partition of $S$ into $m$ non-empty base sets such that the set of their corresponding base vectors $\Vec{v_1}, \dots \Vec{v_m}$ is sparse and the instances in $S$ are for $n>2$ players.
\end{definition}

Observe that the process above also construct an $m$-divisible set. Each iteration $j$ of the process yields a base set $S_j$ (Definition~\ref{base_set}) with base vector $v_j$ and the union of $m$ iterations is an $m$-divisible set $S$ with respect to $S_1, \dots, S_m$ since the set of $\vec{v_1}, \dots, \vec{v_m}$ is sparse (Definition~\ref{sparsity}).

As we saw, partial monotone allocation functions and $m$-divisible sets are closely related objects. We later state Theorem~\ref{construction_thm} and prove it (Section~\ref{proof_construction}) using the terminology of $m$-divisible sets.

\subsection{The Main Technical Theorem} \label{construction_sec}

In this section we state our main technical theorem (Theorem \ref{construction_thm}). This theorem is a construction; it gets as input an $m$-divisible set $S$ (with some parameters) and construct a distribution $D_S$ that is interim IR  $f$-rigid with respect to $S$ (see Definition~\ref{rigid-def}). Where $f$ is an $\varepsilon$-almost linear revenue disagreement function for $\varepsilon$ that depends on the parameters of $S$. We first require some notations and definitions.

\begin{definition}\label{active_def}
Let $S$ be an $m$-divisible set with respect to $S_1, \dots, S_m$. 
Then, player $i$ is \emph{active} in a set $S_j$ if there is an instance $\vec u$ such that $(\Vec{u},i) \in S_j$. For every set $S_j$, we denote by $A^S_j$ its set of active players. Let $A^S$ the set of all active players, i.e., $ A^S = \bigcup\limits_{j\in [m]} A_j^S$.  
Sometimes it will be easier to consider the subsets in which a certain player is active. For that purpose, let ${\Asetforplayer{i}}^S = \{j \in [m] \, |\, i \in A^S_j\}$.
\end{definition}

When considering an $m$-divisible set $S$ with respect to $S_1, \dots, S_m$ we need some notation for the values of the players in the instances of $S$. 
We denote the value of player $i$ in the base vector $\vec{v_j}$ (of $S_j$) by $v_{i,j}$. Similarly, for an active player $i$ in the subset $S_j$, we denote by $u_{i,j}$ its value in the instance $\vec{u}$ such that $(\vec{u}, i) \in S_j$.

The next two definitions of the parameters of an $m$-divisible set $S$ are used to quantify the bound guaranteed by Theorem~\ref{construction_thm} on the approximation ratio of any interim IR mechanism $m$ with agreement ratio of $x$ with $S$.      


\begin{definition}\label{alpha_param}
Let $S$ be an $m$-divisible set with respect to $S_1, \dots, S_m$. 
For every set $S_j$, let $$\alpha^S_j= \frac{\max\limits_{i \in [n]}\{v_{i,j}\}}{\min\limits_{k \in A^S_j}\{u_{k,j} -v_{k,j}\}}$$ Denote by $\alpha_{\text{avg}}^S$ their average, i.e., $\alpha_{\text{avg}}^S = \frac{1}{m}\sum\limits_{j \in [m]}{\alpha_j^S}$.
\end{definition}

The parameter $\alpha^S_j$, defined for the subset $S_j$, measures the ratio between the highest valuation in the base vector $v_j$ and the minimal gap between the threshold $u_{i,j}$ of an active player $i$ in $S_j$ and this player value in the base set $v_{i,j}$. 


\begin{definition}\label{g_param}
Let $S$  be an $m$-divisible set with respect to  $S_1, \dots, S_m$. For every $i \in A$, arrange by ascending order the values $v_{i,j}$ for sets $S_j$ that $i$ is active in (i.e., $i \in A_j$) and denote them by $y_{i,1}, \dots y_{i,{k_i}}$. Let $$g_i^S = \frac{1}{1- \max\limits_{1 \leq j\leq k_i -1}\{\frac{y_{i,j}}{y_{i,{j+1}}}\}} =
1+ \frac{1}{ \min\limits_{1 \leq j\leq k_i -1}\{\frac{y_{i,{j+1}}}{y_{i,{j}}}\} -1} $$
Denote by $g_{\text{avg}}^S$ the average of $(g_i^S)$ over all active players, i.e., $g_{\text{avg}}^S = \frac{1}{|A^S|}\sum\limits_{i \in A^S}(g_i^S)$.  
\end{definition}

The parameter $g_i^S$, for an active player $i$, is related to the growth rate of player $i$'s values in the base vectors of $S_1, \dots, S_m$. Formally, this parameter equal to $1$ plus $1$ over the minimal growth in the values of player $i$ in the base vectors that corresponds to subsets he is active in (the growth is measured between the sorted values), minus one.

\begin{definition} \label{agreement_ratio}
The \emph{agreement ratio} of a mechanism $M$ with some partial allocation function $f(\cdot)$ is the fraction of the allocations that $f$ is defined for which the allocation of $M$ is identical to the allocation of $f$.
\end{definition}


To simplify the proof we require that the $m$-divisible set $S$ satisfy the uniqueness of thresholds property (Definition~\ref{uniqueness}). 

\begin{definition}[uniqueness of thresholds]\label{uniqueness}
An $m$-divisible  with respects to $S_1, \dots, S_m$ has the \emph{uniqueness of thresholds property} if for every player $i$, his thresholds values $u_{i,j}$ ($j \in \Asetforplayer{i}$) 
are distinct and different from his value in the base vectors (i.e., from the values $v_{i,1}, \dots, v_{i,m}$).
\end{definition}



We are now ready to state our main technical theorem:

\begin{theorem}\label{construction_thm}
Let $S$  be an $m$-divisible set with respect to  $S_1, \dots, S_m$ with parameters $|A^S|,  g_{\text{avg}}^S$ and $\alpha_{\text{avg}}^S$ that satisfy the uniqueness of thresholds assumption. There exists a
distribution $\mathcal{F}_S$ on which the approximation ratio of every dominant strategy incentive compatible, interim IR, and deterministic mechanism with agreement
ratio of at most $x$ with $S$ is at most:

$$
\min \{ \contconstruction_S + \, x, \: 1 \}
$$

where $c_S = \cfrac{ |A^S| \cdot g_{\text{avg}}^S+ m\cdot \alpha_{\text{avg}}^S}{|S|} $, and the size of $\mathcal{F}_S$'s support is at most $5|S| + m+ \sum\limits_{j \in [m]} |A_j|^2$.
\end{theorem}

In Section~\ref{ex-rigid-interim} we apply this theorem on sets for which the expression $c_S$
is very small. Thus, the approximation ratio depends mostly on the agreement ratio. 

\begin{claim}\label{almost-linear-claim}
In Theorem~\ref{construction_thm}, every interim IR, dominant strategy incentive compatible and deterministic mechanism with agreement ratio at most $x \in [1-c_S, 1)$ has approximation ratio less than 1. 
\end{claim}

The proof of Claim~\ref{almost-linear-claim} is deferred to Appendix~\ref{al-sec}. 

\begin{corollary} \label{rigid_construction}
Let $S$ be an $m$-divisible set with parameter $c_S$, then there exists a distribution $\mathcal{F}_S$ that is interim IR $f_S$-rigid with respects to $S$ (Definition~\ref{rigid-def}). Where $f_S$ is $c_S$-almost linear revenue disagreement function (Definition~\ref{al-def}).  
\end{corollary}

\section{Proof of Theorem~\ref{construction_thm}}\label{proof_construction}

Firstly, we need additional notations, observations, and a characterization to prove the theorem (Section~\ref{per-proof}). 
We divide the proof of the theorem into two. The first part (Section~\ref{F_S}) is the construction of the distribution $\mathcal{F}_S$ (from the statement of the theorem). The second part (Section~\ref{analysis_of_revenue}) analyses the revenue that can be extracted from this distribution $\mathcal{F}_S$ in terms of the agreement ratio. The proof of Theorem~\ref{construction_thm} considers a specific $m$-divisible set $S$. Thus, to simplify notation, we drop the superscript $S$ from all of parameters. 

\subsection{Preliminaries for the Proof} \label{per-proof}

We rely on the following characterization result of Feldman and Lavi for deterministic dominant strategy incentive compatible and interim mechanisms.  
\begin{proposition} [Feldman and Lavi \cite{FL-thesis}] \label{characterization} 
Fix a deterministic dominant strategy incentive compatible and interim IR auction A. Then, there exists a deterministic dominant strategy incentive compatible, normalized and ex-post IR auction $A^*$ and fees $c_i: D_{-i} \to \mathbb{R}$ such that for any instance ($v$) the winner in $A^*$ is the same as in $A$ and his payment in $A$ is equal to his payment in $A^*$ plus the relevant fees (i.e., $p_i^A(v)= p_i^{A^*}(v) + c_i(v_{-i})$).  
Furthermore, since $A$ is an interim IR mechanism, the fees must satisfy:

\begin{equation} \label{characterizarion_condition}
    \mathop{\mathbb{E}}\nolimits_{{v}_{-i}\sim \mathcal{F}_{-i} | v_i=v_i}[c_i(v_{-i})]  \leq \pi_i^{A^*(D)}(v_i) \quad \quad  \forall i \in [n], \, \forall v_i \in D_i
\end{equation}

where $\pi_i^{A^*(\mathcal{F})}(v_i)$ is the expected profit of player $i$  w.r.t. $A^*$ conditioned on his value being $v_i$.
\end{proposition}

Note that the fees charged from player $i$ are not a function of his own value, similarly to the payment. 


\begin{definition}
The conditional probability matrix of bidder $i$ w.r.t a discrete distribution $\mathcal{F}$ over a domain $D$ is a matrix $CP_i(\mathcal{F})$ of dimensions $|D_i| \times |D_{-i}|$ where for every $ 1 \leq k \leq |D_i|$ and every $1 \leq j \leq |D_{-i}|$ we have: 
$$
[CP_i(\mathcal{F})]_{(k,j)} = \Pr_{v \sim F}(v_{-i}=v^j_{-i}| v_i=v_i^k)
$$ 
\end{definition}

$[CP_i(\mathcal{F})]_{(v_i)}$ is the row in the conditional probability matrix $CP_i(\mathcal{F})$ that corresponds to the value $v_i \in D_i$. 

The fees described in Proposition~\ref{characterization} are functions. However, we sometimes refer to them as a column vector $\Vec{c_i}$ of dimension $D_{-i}$ where the $j$'th entry corresponds to the fess charged from player $i$ when $v_{-i}$ equals the $j$'th value in $D_{-i}$. In this case the condition in~\eqref{characterizarion_condition} can be rewritten as:

\begin{equation} \label{chr_cond_matrix}
    [CP_i(\mathcal{F})\cdot {\Vec{c_i}}]_k  \leq \pi_i^{A^*(D)}(v_i^k) \quad \quad  \forall i \in [n], \, \forall k=1, \dots, |D_i|
\end{equation}

By Proposition~\ref{characterization}, we have that every deterministic, dominant strategy incentive compatible, and interim IR mechanism $M$ can be described by a deterministic, dominant strategy incentive compatible and ex-post IR mechanism $B$ and vector of fees $\Vec{c_i}$, one for every player $i$ (recall that $A$ has the same allocation function as $M$ and that $M$'s revenue equals to the sum of $A$'s revenue and the expected sum of fees $\Vec{c_i}$ from all players). If we are given $B, \Vec{c_i}$ as the description of $M$, we say that $M$ is in the \emph{standard form}.


\begin{observation} \label{efs_bounded_by_value}
Let $B, \Vec{c_i}$ be a deterministic,  interim IR and dominant strategy incentive compatible mechanism in the standard form. Then, for every player $i \in [n]$ and every value $v_i \in D_i$ it holds that $[CP_i(\mathcal{F})]_{(v_i)}\cdot \Vec{c_i} \leq v_i$.
\end{observation}

\begin{observation} \label{efs_upper_bound}
Let $\mathcal{F}$ be a joint distribution over the values of $n$ bidders ($D$) and let $(B, \Vec{c_i})$ be a deterministic, interim IR and dominant strategy incentive compatible mechanism in the standard form. Let $i$ be a player with value $v_i \in D_i$ for which exist non negative numbers $\alpha_{1}, \dots \alpha_{|D_i|}$ such that
\begin{align}
    [CP_i(\mathcal{F})]_{(v_i)} = \sum\limits_{v_i^j \in D_i}{\alpha_{j}\cdot [CP_i(\mathcal{F})]_{(v^j_i)}}
\end{align}
Then:
\begin{equation}
    [CP_i(\mathcal{F})]_{(v_i)}\cdot \Vec{c_i} \leq \sum\limits_{v_i^j \in D_i}{\alpha_{j}\cdot \pi_i^{B(\mathcal{F})}(v_i^j)} \leq \sum\limits_{v_i^j \in D_i}{\alpha_{j}\cdot v_i^j}
\end{equation}
\end{observation}

Note that the condition over $v_i$ in Observation~\ref{efs_upper_bound} is actually that the row vector corresponding to $v_i$ in $[CP_i(\mathcal{F})]$ is a convex combination of the other row vectors in $[CP_i(\mathcal{F})]$.

\begin{proof}[Proof of Observation~\ref{efs_upper_bound}]
Recall the constraints on the fees described in~\eqref{chr_cond_matrix}, then:
$$
[CP_i(\mathcal{F})]_{(v_i)}\cdot \Vec{c_i}= \sum\limits_{v_i^j \in D_i}{\alpha_{j}\cdot [CP_i(D)]_{(v^j_i)}\cdot \Vec{c_i}} \; \underbrace{\leq}_{\alpha_j \geq 0 \atop \eqref{chr_cond_matrix}} \:\, \sum\limits_{v_i^j \in D_i}{\alpha_{j}\cdot \pi_i^{B(\mathcal{F})}(v_i^j)} \underbrace{\leq}_{~\ref{efs_bounded_by_value}} \: \sum\limits_{v_i^j \in D_i}{\alpha_{j}\cdot v^j_i}
$$
\end{proof}

\subsection{Part 1: the Construction}\label{F_S}

The distribution $\mathcal{F}_S$ is composed of four subdistributions: \emph{$P,E,O,R$} (a subdistribution is a restriction of the original distribution function to some subset of its support). 

The subdistribution $P$ (Section~\ref{P}) consists of $\sizeActive$ (the number of active players in $S$) subdistributions. Each such subdistribution is an equal-revenue distribution with respect to a different active player (Definition~\ref{equal-revenue-property}). The revenue that any mechanism can extract from this subdistribution is low comparing to the expected social welfare. Moreover, in the optimal \emph{ex-post IR} mechanism, the \emph{profit} of the active player is very high, in fact close to its value. 

Most of the revenue that can be extracted in $\mathcal F_S$ will be from instances in the support of the subdistribution $E$ (Section~\ref{E}), in the form of fees. The subdistribution $O$ (Section~\ref{O}) is based on the \emph{original} set $S$.  
We use the subdistribution $R$ (Section~\ref{R}) to \emph{restrict} how fees can be extracted. This will be useful for the approximation claim. We now define some constants that are used in the construction.

\begin{itemize}
    \item Let $0 < \epsilonvalintermidate < 0.1$ be a small constant that can be arbitrarily small. 
    \item Let $0 < \epsilonvaldevi< \min \{\, 2 \cdot\min\limits_{j \in [m], i \in A_j}\{u_{i,j}-v_{i,j}\},\; \epsilonvalintermidate\}$ be small enough constant that can be chosen to be arbitrarily small.
    \item Set $\basevectorprob{} = 0.1$. 
    \item By renaming and without loss of generality, we assume that the first $a$ players are the active players.
    \item For every player $i$ and every $j \in \Asetforplayer{i}$, obtain $\thsintermediate{i,j}$ by decreasing the value of ${u}_{i,j}$ (defined in Section~\ref{construction_sec}) by at most $\epsilonvaldevi$ in order to make sure that these values are different from each other and from every value $v_{i,k}$ (defined in Section~\ref{construction_sec}), for each $v_{i,k}$ from the sparse set of base vectors.
\end{itemize}

\subsubsection{The Subdistribution $P$}\label{P}

We now construct the subdistribution $P$ that consists of $\sizeActive$ equal-revenue subdistributions, one for each active player.

\begin{definition}\label{equal-revenue-property}
We say that a subdistribution $F$ over the values of $n$ bidders has the \emph{equal revenue property} with respect to player $i$ if the following two conditions hold:
\begin{enumerate}[label=(\arabic*)]
    \item\label{same_v_-i} All vectors in the support of the subdistribution have the same $v_{-i}$.
    \item For any two values $v,v'$ in the support of player $i$ in $F$ it holds that $\Pr_{v_i\sim F}[v_i\geq v]\cdot v=\Pr_{v_i\sim F}[v_i\geq v']\cdot v'$. \label{same_rev}
\end{enumerate}
\end{definition}

For every active player $i \in A$, we construct a subdistribution with the equal revenue property with the following properties. In every instance $\vec t$ in this subdistribution, $t_{-i}$ is the same (Condition \ref{same_v_-i}), and the values in $t_{-i}$ are very small relative to each of player $i$'s values in this subdistribution. Also, almost all of the support of player $i$'s in this subdistribution consists of player $i$'s values in the base vectors that he is active in (i.e., the values $v_{i,j}$ for $j \in \Asetforplayer{i})$). We choose the probabilities in this subdistribution so that Condition \ref{same_rev} is satisfied.

The rest of this subsection is essentially devoted to showing that there are probabilities that result in an equal-revenue distribution.

For each player $i$, the values in the corresponding subdistribution are denoted by $y_{i,0}, \dots, y_{i,k_i +1}$ (where $k_i$ is the number of subsets that player $i$ is active in, $k_i = |\Asetforplayer{i}|$). Most of those values are player $i$'s values in the base vectors that he is active in. To make sure that every point in $S$ will have the same contribution to the revenue from fees, we add a value $y_{i,0}$ to the support of player $i$ (see Equation~\ref{minimal_efs_equal}). The additional role of the value $y_{i,0}$ is to control the social welfare of the constructed subdistribution ${P}$\footnote{When the valuations of the other players are much smaller, the optimal ex-post IR revenue of a subdistribution with the equal revenue property with respect to player $i$ equals to the lowest value in player $i$'s support.} and thus we choose them to be small enough so we can extract almost all of the social welfare of $P$ by fees (Equation~\ref{efs_constraint}).

For every $i \in A$, arrange in ascending order the values $v_{i,j}$ for $j \in \Asetforplayer{i}$ and denote them by $y_{i,1} < \dots < y_{i,k_i}$. Define the functions:

$$\sigma_i:\Asetforplayer{i} \to [k_i] \text{ s.t.  } \sigma_i(j)= r \text{ where } y_{i,r}= v_{i,j}$$
$$\sigma_i^{-1}: [k_i] \to \Asetforplayer{i} \text{ is } \sigma_i\text{'s} \text{ inverse}$$

and let:
$$ d_i = 1-\max\limits_{1 \leq j\leq k_i -1}\{\frac{y_{i,j}}{y_{i,j+1}}\}. $$
Let $y_{1,0}, \dots y_{\sizeActive,0}$ be a positive solution to the equations in \eqref{minimal_efs_equal} that satisfies the constraints in \eqref{delta_constraint}  

\begin{equation}\label{minimal_efs_equal}
    y_{1,0} \cdot d_1 = y_{2,0} \cdot d_2 = \dots = y_{\sizeActive,0} \cdot d_{\sizeActive}
\end{equation}
\begin{subequations}\label{delta_constraint}
\begin{equation}\label{efs_constraint}
    \frac{1}{m}(\basevectorprob{} -\epsilonvallowest) \geq  \max\limits_{i \in  \Asetforsubset{j} } \{\cfrac{(1-\basevectorprob{})\cdot(y_{i,0}\cdot d_i)}{\sizeActive\cdot (\thsintermediate{i,j} - v_{i,j})}\}  \quad   \forall j \in [m]\\
\end{equation}
\begin{equation}\label{y1>y0}
     \quad \qquad \qquad  y_{i,0} <  y_{i,1}  \quad \quad \quad  \: \ \forall i \leq \sizeActive     
\end{equation}
\end{subequations}

\begin{lemma} \label{lemma:y_0_comp}
There exist positive numbers $y_{1,0}, \dots y_{\sizeActive,0}$ that are the solutions to  \eqref{minimal_efs_equal} and satisfy the constraints in \eqref{delta_constraint}.
\end{lemma}

The proof of Lemma~\ref{lemma:y_0_comp} is deferred to Appendix~\ref{proof-of-4.2}.
Denote by $ e =y_{1,0}\cdot d_1$ (recall the the LHS of this definition is equal for every active player, Equation~\ref{minimal_efs_equal}) and by $y_{i,k_i+1} = \frac{y_{i,k_i}}{1-d_i}$ for every active player $i$.

Next, we compute probabilities $q_{i,0}, \dots q_{i,k_i+1}$ for the points $y_{i,0}, \dots y_{i,k_i+1}$ such that they will constitute a subdistribution with the equal revenue property for player $i$ (this will satisfy condition~\ref{same_rev} in Definition~\ref{equal-revenue-property}). Formally, we find probabilities for the values in player $i$'s support such that by offering player $i$ a price equal to each such value, the expected revenue is the same (this is the role of Equation~\ref{EQD}). 

\begin{align}\label{EQD}
\begin{split}
    y_{i,0} & = y_{i,j}(q_{i,j} + \dots q_{i,k_i+1}) \quad \forall \, 1 \leq j \leq k_i+1 \\
    q_{i,0} &= 1 - (q_{i,1} + \dots q_{i,k_i+1})
\end{split}
\end{align}

\begin{lemma} \label{lemma:EQD_probabilites}
Equation set~\eqref{EQD} has a solution that yields a probability distribution. 
\end{lemma}

The proof of Lemma~\ref{lemma:EQD_probabilites} is deferred to Appendix~\ref{proof-of-4.3}.  Let  $0 < \epsilonvallowest < \epsilonvalintermidate $ be a small enough constant that satisfies:
    \begin{equation*}
        \epsilonvallowest < \min \{\frac{\epsilonvalintermidate}{ 5\cdot n\cdot m + 3n +n^2 +n^2\cdot m},\,  \min\limits_{i \in [n] \atop j \in [m]} \{ v_{i,j}\}, \, \min{i \in A} \{y_{i,0}\}\}.
    \end{equation*}
Finally, we construct the subdistribution. 
We find an arbitrarily small value $ \epsilonvaleqr{i}$ for the valuations of the other players in the subdistribution of player $i$. 
For every $i \in A$, let $0 < \epsilonvaleqr{i} < \epsilonvallowest$ be a small constant such that all these constants are different from each other, every ${v_{i,j}}$ (for every $S_j$ and every player $i$), every $u_{i,j}$ or $\thsintermediate{i,j}$ (for every $S_j$ and every $i \in A_j$).
Now, for every active player $i$, we add instances of the form $(y_{i,j}, v_{-i})$ (for every $ 1\leq j\leq k_i+1$) to his subdistribution with probabilities $q_{i,j}\cdot \frac{(1-\basevectorprob{})}{\sizeActive}$, where $v_{-i} = (\epsilonvaleqr{i}, \dots, \epsilonvaleqr{i})$. 

Observe that for every player $i$ we get a subdistribution with the equal revenue property with respect to him. Condition \ref{same_v_-i} of Definition \ref{equal-revenue-property} is met as we constructed every instance with the same value of $v_{-i}$. Condition \ref{same_rev} is met as we constructed the probabilities by making sure it is satisfied (Equation~\ref{EQD}).

We finish this section with stating some bound that we use in the analysis part of the proof (Section~\ref{analysis_of_revenue}). The proof of Lemma~\ref{lemma:expected_sw_bound} is deferred to Section~\ref{proof-of-4.4}.


\begin{lemma} \label{lemma:expected_sw_bound}
For every $i\in A, 1\leq j \leq k_i$, it holds that $e = y_{i,0}\cdot d_i \leq y_{i,j} \cdot q_{i,j}$. 
\end{lemma}

\subsubsection{The Subdistribution $E$}\label{E}

This subdistribution consists of the base vectors $\Vec{v_1}, \dots \Vec{v_m}$ (Definition~\ref{base_set}) of the subsets $S_1, \dots, S_m$. Most of the revenue of the optimal mechanism is from fees charged in these instances. We want these instances to have small probabilities so their affect on the social welfare of $\mathcal{F}_S$ and thus their affect on the ex-post revenue that can be extracted from $\mathcal{F}_S$ will be small. This is the role of Equation~\ref{prob_base_vector}. 

We compute the probability $\basevectorprob{j}$ of each base vector $\Vec{v_j}$ in $E$. For every subset $S_j$, define:

\begin{equation} \label{prob_base_vector}
    \basevectorprob{j} =  \max\limits_{i \in  \Asetforsubset{j} } \{\cfrac{(1-\basevectorprob{})\cdot e}{\sizeActive\cdot (\thsintermediate{i,j} - v_{i,j})}\} = \cfrac{(1-\basevectorprob{})\cdot e}{\sizeActive\cdot \min\limits_{i \in \Asetforsubset{j}}\{\thsintermediate{i,j}-v_{i,j}\}}.
\end{equation}

Observe that inequality~\eqref{efs_constraint} implies that $0 < \basevectorprob{j} < 1$ and that $\epsilonvallowest + \sum_{j \in [m]}{\basevectorprob{j}} \leq \basevectorprob{}-\epsilonvallowest = 0.1$.
For every $S_j$ we add the point  $v_j$ with probability $\basevectorprob{j}$ to $E$.

\subsubsection{The Subdistribution $O$}\label{O}

This subdistribution is based on the instances $(u_{i,j}, (\Vec{v_j})_{-i})$ of the $m$-divisible set $S$.
Each of its instances $(u_{i,j}, (\Vec{v_j})_{-i})$ is added to the distribution together with a similar instance $(u'_{i,j}, (\Vec{v_j})_{-i})$ that is defined next. Both instances are given arbitrarily small probabilities.
The purpose of the first instance  $(u_{i,j}, (\Vec{v_j})_{-i})$ is to condition the extraction of fees of some mechanism $M$ from player $i$ in the instance $\vec{v_j}$ on its agreement with $S$ on the allocation in the instance $(u_{i,j}, (\Vec{v_j})_{-i})$. We use the second instance $(u'_{i,j}, (\Vec{v_j})_{-i})$ to limit the amount of fees that can be charged from player $i$ in the instance $\vec{v_j}$. 

\begin{observation} \label{relation_maintained}
For every $S_j$ and every $i \in A_j$ it holds that:
\begin{subequations}
\begin{equation}
    v_{i,j} +\min\limits_{k \in A_j}\{\thsintermediate{k,j}-{v}_{k,j}\} \leq u_{i,j}
\end{equation}
\begin{equation}
    v_{i,j} +\min\limits_{k \in A_j}\{\thsintermediate{k,j}-{v}_{k,j}\} - \epsilonvaldevi > v_{i,j}
\end{equation}
\end{subequations}
\end{observation}

\begin{proof}[Proof of Observation~\ref{relation_maintained}]
\begin{subequations}
\begin{equation}
    v_{i,j} +\min\limits_{k \in A_j}\{\thsintermediate{k,j}-{v}_{k,j}\} \leq v_{i,j} + \thsintermediate{i,j}-{v}_{i,j} = \thsintermediate{i,j} \leq u_{i,j}
\end{equation}
\begin{equation}
    v_{i,j} +\min\limits_{k \in A_j}\{\thsintermediate{k,j}-{v}_{k,j}\} -\epsilonvaldevi \underbrace{>}_{\thsintermediate{k,j} \geq u_{k,j}- \epsilonvaldevi \atop 2\epsilonvaldevi < u_{k,j}- v_{k,j}} v_{i,j} 
\end{equation}
\end{subequations}
\end{proof}

For every subset $S_j$ and every player $i\in \Asetforsubset{j}$, let $u'_{i,j}= v_{i,j} + \min\limits_{k \in \Asetforsubset{j}}\{\thsintermediate{k,j}-v_{k,j}\}-\epsilon'$, by arbitrarily small $\epsilon' < \epsilonvaldevi$ (that might be different for each $i,j,k$ such that these values are different from each other and from every value $v_{i,k}$, for each $v_{i,k}$ from the sparse set of base vectors. 

From Observation~\ref{relation_maintained} we have that $u_{i,j}\geq {u'}_{i,j} > v_{i,j} $. The instances in the support of this subdistribution will be have small probabilities. Let $0 < \epsilonprob{} < \epsilonvallowest$  be a small enough constant that satisfies both:
\begin{subequations}
\begin{equation}
    \epsilonprob{} \cdot (2n\cdot m) \cdot \max\limits_{j \in [m], i \in A_j}\{u_{i,j}\} + \epsilonvallowest \cdot \epsilonprob{} \cdot (n+n\cdot m + n^2 \cdot m) < \epsilonvalintermidate.
\end{equation}
\begin{equation} \label{smaller_than_lowest}
    \epsilonprob{}   < \min\{\frac{\epsilonvallowest}{\max\limits_{j \in [m], i \in A_j}\{u_{i,j}\}}, \frac{\epsilonvallowest}{4n^2m}\}.
\end{equation}
\end{subequations}

The subdistribution $O$ is obtained by including the 
instances $({u'}_{i,j}, (v_j)_{-i})$ and $({u}_{i,j}, (v_j)_{-i})$ and assigning them both\footnote{If $u_{i,j} = u'_{i,j}$ we get the same point and we assign it a probability of $\epsilonprob{}$.} probability of $\epsilonprob{}$.

\subsubsection{The Subdistribution $R$}\label{R}

Instances in this subdistribution are used to restrict the fees that can be charged from some player $i \in [n]$. The probabilities of all instances in the support of this subdistribution will be very small and the values of player $i$ in the support will be very small as well.

For every player $i$ and every value he gets in the subdistributions $P, E, O$ that is large enough $v_i > \epsilonvallowest$, we go over almost (we exclude the values $v_{-i} = \Vec{(v_j)}_{-i}$ for subset $S_j$ that player $i$ is active in) all of the values $v_{-i}$ such that $(v_i, v_{-i})$ is an instance in one of the subdistributions $P, E, O$ and make sure that no fees can be extracted from player $i$ when the valuations of the other players are $v_{-i}$.
We do that by adding an instance in which player $i$ has arbitrarily small value (smaller than $\epsilonvallowest$) and the other players have valuations $v_{-i}$, we assign this instance arbitrarily small probability:


\begin{itemize}
    \item For every active player $i \leq \sizeActive$: \begin{itemize}
        \item Let $0 < \epsilonvalrestrictefs{i} < \epsilonvallowest$ be a small constant such that all these constants are different from each other and every other value in the support of $\mathcal{F}_S$ so far.
        \item Assign the point $(v_i, v_{-i})=(\epsilonvalrestrictefs{i},\epsilonvaleqr{i} \dots \epsilonvaleqr{i})$  probability  $\epsilonprob{}$ and add it to $R$. 
    \end{itemize}
    \item For every subset $S_j$, every two different active players in it $i \neq k \in \Asetforsubset{j}$: \begin{itemize}
        \item Let $0 < \epsilonvalrestrictefs{i,j,k}, \epsilonvalrestrictefs{i,j,k}' < \epsilonvallowest$ be small constants such that all these constants are different from each other and every other value in the support of $\mathcal{F}_S$ so far.
        \item Assign the point $(v_i,v_k,v_{-\{i,k\}} )=(\epsilonvalrestrictefs{i,j,k},{{u'}_{k,j}}, (v_j)_{-\{i,k\}})$ probability $\epsilonprob{}$ and add it to $R$. 
        \item Assign the point $(v_i,v_k,v_{-\{i,k\}} )=(\epsilonvalrestrictefs{i,j,k}',{{u}_{k,j}}, (v_j)_{-\{i,k\}})$ probability $\epsilonprob{}$ and add it to $R$. 
    \end{itemize}        
\end{itemize}

Observe that the claim about the support size of $F_S$ follows by summing together the number of instances in the subdistributions $P,E,O$ and $R$.

\subsection{Part 2: Analysis of the Revenue of the Distribution $\mathcal{F}_S$} \label{analysis_of_revenue}

In this section we analyze the revenue that can be extracted in the distribution $\mathcal{F}_S$. We derive Theorem \ref{construction_thm} by the following two propositions:

\begin{proposition} \label{lower_bound_on_revenue}
There exists a deterministic, dominant strategy incentive compatible, and interim IR mechanism which extracts in expectation over $\mathcal F_S$ revenue of at least:  $$ \sum\limits_{j=1}^m \sum\limits_{i \in \Asetforsubset{j}} e\cdot \frac{1-\delta}{\sizeActive}$$ 
\end{proposition}

\begin{proposition}\label{upper_bound_on_revenue}
Let $M$ be a deterministic, dominant strategy incentive compatible, and interim IR mechanism with agreement ratio of $x$ with $S$. The expected revenue of $M$ over $\mathcal{F}_S$ is at most:
$$
\sum\limits_{i=1}^{\sizeActive}{y_{i,0} \cdot \frac{1-\basevectorprob{}}{\sizeActive}} + \sum\limits_{j=1}^{m} \basevectorprob{j}\cdot \max\limits_{i \in {n}}\{v_{i,j}\} + \frac{1-\basevectorprob{}}{\sizeActive}\cdot e \cdot x\cdot|S|
$$
\end{proposition}

\begin{proof}[Proof of Theorem~\ref{construction_thm}]

We apply both Proposition ~\ref{lower_bound_on_revenue} and Proposition~\ref{upper_bound_on_revenue} to get that the approximation ratio of every deterministic, dominant strategy incentive compatible and interim IR mechanism $M$ with agreement ratio of $x$ with $S$ is at most:

\begin{align*}
\frac{\mathop{\mathbb{E}}\limits_{v\sim F}[REV_M(v)]}{\mathop{\mathbb{E}}\limits_{v\sim F}[REV_{opt}(v)]}  \leq & \,
\frac{ \sum\limits_{i=1}^{\sizeActive}{y_{i,0} \cdot \frac{1-\basevectorprob{}}{\sizeActive}} + \sum\limits_{j=1}^{m}{\basevectorprob{j}\cdot \max\limits_{i \in {n}}\{v_{i,j}\}} + \frac{1-\basevectorprob{}}{\sizeActive}\cdot e  \cdot x\cdot|S|}{\sum\limits_{j=1}^m \sum\limits_{i \in \Asetforsubset{j}} e\cdot \frac{1-\basevectorprob{}}{\sizeActive}} =
 \frac{ \sum\limits_{i=1}^{\sizeActive}{\frac{1}{d_i}} + \sum\limits_{j=1}^{m}{\frac{\max\limits_{i \in {n}}\{v_{i,j}\}}{\min\limits_{i \in \Asetforsubset{j}}\{\thsintermediate{i,j}-v_{i,j}\}}} +  x\cdot|S|}{\sum\limits_{j=1}^m \sum\limits_{i \in \Asetforsubset{j}} 1}\\
 \leq & \, \frac{ \sum\limits_{i=1}^{\sizeActive}{ \frac{1}{d_i}} + \sum\limits_{j=1}^{m}{\frac{\max\limits_{i \in {n}}\{v_{i,j}\}}{\min\limits_{i \in \Asetforsubset{j}}\{{u}_{i,j}-v_{i,j}-\epsilonvaldevi\}}} +  x\cdot|S|}{\sum\limits_{j=1}^m \sum\limits_{i \in \Asetforsubset{j}} 1} =  \frac{ \sum\limits_{i=1}^{\sizeActive}{g_i} + \sum\limits_{j=1}^{m}{\alpha_j}}{|S|} +x
=  \cfrac{  |A| \cdot g_{\text{avg}}+ m\cdot \alpha_{\text{avg}}}{|S|} + \, x
\end{align*}

The first equality is because $y_{i,0} = \frac{e}{d_i}$ and $\basevectorprob{j} =  \max\limits_{i \in \Asetforsubset{j}}\{{\cfrac{(1-\basevectorprob{})\cdot(y_{i,0}\cdot d_i)}{\sizeActive\cdot (\thsintermediate{i,j} -v_{i,j})}}\}
    =\frac{1-\basevectorprob{}}{\sizeActive}\cdot \cfrac{e}{\min\limits_{i \in \Asetforsubset{j}}\{\thsintermediate{i,j}-v_{i,j}\}}$. The second inequality is because  ${\thsintermediate{i,j} \geq u_{i,j} -\epsilonvaldevi}$. 
\end{proof}


\subsection{Proof of Proposition~\ref{lower_bound_on_revenue}} \label{lower_bound_sec}


We prove this proposition by providing an ex-post IR, deterministic and dominant strategy incentive compatible mechanism $\expostmechanism'$ with fees $\vec{c'_i}$ that together compose an interim IR mechanism $M'$ with the required revenue.   

We start with describing the allocation function of $\expostmechanism'$ and its payments in the instances of $\mathcal F_S$ by specifying its thresholds. Recall that in an ex-post IR mechanism the payment of a winning bidder equals his threshold and that a losing bidder pays $0$. 

For every active player $i \in A$ and every subset $S_j$ that he is active in $j \in \Asetforplayer{i}$, let player $i$'s threshold for $v_{-i} = \Vec{(v_j)}_{-i}$ be $v_{i,j} + \epsilon$ for some arbitrarily small value of $\epsilon > 0$. 
For every active player $i \in A$, let player $i$'s threshold for $v_{-i} = (\epsilonvalrestrictefs{i},\dots,\epsilonvalrestrictefs{i})$ be $0$.
All other thresholds are set to $\infty$, i.e., $\expostmechanism'$ does not allocate the item in any other case. 

\begin{claim} \label{legal-ex-post}
$\expostmechanism'$ is a deterministic, ex-post IR , and dominant strategy incentive compatible mechanism.
\end{claim}

\begin{proof}
$\expostmechanism'$ is clearly a deterministic, ex-post IR mechanism. It is also a dominant strategy mechanism since each player is allocated the item if his value is more than some threshold that does not depend on his value. It remains to show that the mechanism is feasible, i.e., the item is not allocated to two players in the same instance.

Consider some instance $\vec{v}$. If $\expostmechanism'$ allocates the item to some player $i$ in $\vec v$, then when the values of the other players are $\vec{v}_{-i}$, the threshold of player $i$ is some $t_i \leq \vec{v}_i$.

Observe that for every instance $\vec{v}$ in the support of the subdistribution $P$,  $\expostmechanism'$ only 
allocates the item to the $i$'th player. For every other player $j \neq i$ the values of $v_{-j}$ of the instances in this subdistribution are unique. Thus, the threshold of every $j \neq i$ is $\infty$. 

The other case we need to consider is that $v_{-1} = (\vec{v_j})_{-1}$ and $v_{-2} = (\vec{v_k})_{-2}$ for some $j,k \in [m]$. However, since the base vectors $\vec{v_1}, \dots, \vec{v_m}$ are sparse (Definition~\ref{sparsity}) it must be the case that $j=k$ and thus $\vec{v} = \vec{v_j}$. Since the threshold of every player $i$ is larger than his value in $\vec{v_j}$, it cannot be that both players $1$ and $2$ are to be allocated the item in $\vec{v_j}$.
\end{proof}

We set values for the fees $\vec{c'_i}$ and prove that $\expostmechanism'$ together with $\vec{c'_i}$ is an interim IR, dominant strategy incentive compatible and deterministic mechanism (Lemma~\ref{lemma:legal-interim}). For every active player $i \in A$ and every subset $S_j$ that he is active in we set $c'_i((v_j)_{-i}) = \frac{\frac{1-\basevectorprob{}}{\sizeActive}\cdot e}{\basevectorprob{j}}$. For every other value of $v_{-i}$ we set $c'_i(v_{-i})=0$. 

\begin{lemma} \label{lemma:legal-interim}
$\expostmechanism'$ with fees charged according to $\vec{c'_i}$ is an interim IR, dominant strategy incentive compatible and deterministic mechanism.
\end{lemma}

The proof of Lemma~\ref{lemma:legal-interim} is deferred to Appendix~\ref{proof-of-4.11}.

The expected revenue of the interim IR mechanism ($\expostmechanism'$, $c'_i$)  with respect to $\mathcal F_S$ is:
$$
\sum\limits_{j=1}^m \sum\limits_{i \in \Asetforsubset{j}} e\cdot \frac{1-\basevectorprob{}}{\sizeActive}
$$

\subsection{Proof of Proposition~\ref{upper_bound_on_revenue}}\label{upper_bound_sec}

In this section we prove an upper bound on the revenue of a deterministic, dominant strategy incentive compatible and interim IR mechanism $M$ with an agreement ratio of $x$ with $S$. 
Let $\expostmechanism, \Vec{c_i}$ be $M$'s standard form guaranteed by Proposition~\ref{characterization}. Recall that $\expostmechanism$ is an ex-post IR, deterministic and dominant strategy incentive compatible mechanism with the same allocation function of $M$ and thus the same agreement ratio of $x$ with $S$, and the $\Vec{c_i}$'s are vectors of fees taken from every player $i$ such that they satisfy condition~\ref{chr_cond_matrix}. The reason we can use this characterization is that $M$'s revenue is equal to the sum of $\expostmechanism$'s revenue and the revenue extracted by the $\vec{c_i}$'s.

We define some notation that is used in the analysis the two main claims (Claim~\ref{ex-post-revenue} and Claim~\ref{efs-revenue}) used to prove this proposition. 
For every player $i \in [n]$ and every subset $S_j$, let $\thsbasevec{i,j}$ be the threshold that $\expostmechanism$ assigns player $i$ when $v_{-i} = \Vec{(v_j)}_{-i} $. Let $\thsindicbasevec{i,j}$ be an indicator variable that equals 1 if $\thsbasevec{i,j} < v_{i,j}$ and 0 otherwise.

For every active player $i \in [\sizeActive] $, let $\thserd{i}$ be the threshold that $\expostmechanism$ assigns player $i$ when $v_{-i} = (\epsilonvalrestrictefs{i}\dots \epsilonvalrestrictefs{i}) $. For every active player $i \in A$ and every $j \in \Asetforplayer{i}$ let $\feemech{i,j} = \min \{ (y_{i,\sigma_i(j)}-\thserd{i})\cdot q_{i,\sigma_i(j)}, \, e\}$ and let $\chi_{i,j}$ be an indicator variable that equals $1$ if $\expostmechanism$ agrees with $S$ on the allocation in the instance $({u}_{i,j}, (v_j)_{-i})$ and $0$ otherwise.

We will prove a tighter bound on the revenue that will be useful for proving Claim~\ref{almost-linear-claim}. 

\begin{proposition}\label{tigher-ub}
The expected revenue of $M$ over $\mathcal{F}_S$ is at most:
$$
\sum\limits_{i=1}^{\sizeActive}{\thserd{i}}\cdot (\sum\limits_{\substack{j \in \Asetforplayer{i} \\ s.t. y_{i,j}\geq \thserd{i} }} {q_{i,\sigma_i(j)}\cdot \frac{1-\basevectorprob{}}{\sizeActive}}) + \sum\limits_{j=1}^{m} \basevectorprob{j}\cdot \max\limits_{i \in {n}}\{v_{i,j}\} +  \sum\limits_{j=1}^m \sum\limits_{i \in \Asetforsubset{j}} \feemech{i,j}\cdot \frac{1-\delta}{\sizeActive} \cdot \chi_{i,j}
$$
\end{proposition}

Observe that proving this proposition is enough as $\feemech{i,j} \leq e$ for every active player $i$ and every $j \in \Asetforplayer{i}$, and since  $$\thserd{i}\cdot (\sum\limits_{\substack{j \in \Asetforplayer{i} \\ s.t. y_{i,j}\geq \thserd{i} }} {q_{i,\sigma_i(j)}\cdot \frac{1-\basevectorprob{}}{\sizeActive}})$$ is the ex-post revenue mechanism $M$ extracts from the equal revenue distribution of player $i$ and thus it holds that :
$$\thserd{i}\cdot (\sum\limits_{\substack{j \in \Asetforplayer{i} \\ s.t. y_{i,j}\geq \thserd{i} }} {q_{i,\sigma_i(j)}\cdot \frac{1-\basevectorprob{}}{\sizeActive}}) \leq y_{i,0} \cdot \frac{1-\basevectorprob{}}{\sizeActive}.$$

We analyze separately the revenue an ex-post IR mechanism $\expostmechanism$ can extract (Claim~\ref{ex-post-revenue}) and the revenue that can be extracted using fees $\vec{c_i}$ (Claim~\ref{efs-revenue}). Proving these two claims will conclude the proof of Proposition~\ref{tigher-ub}.

\begin{claim} \label{ex-post-revenue}
The extracted revenue in $\mathcal{F}_S$ by the deterministic, dominant strategy incentive compatible and ex-post IR mechanism $B$ is at most:
$$
REV(\expostmechanism,F) \leq \sum\limits_{i=1}^{\sizeActive}{\thserd{i}}\cdot (\sum\limits_{\substack{j \in \Asetforplayer{i} \\ s.t. y_{i,j}\geq \thserd{i} }} {q_{i,\sigma_i(j)}\cdot \frac{1-\basevectorprob{}}{\sizeActive}})  + \sum\limits_{j \in [m]}  \basevectorprob{j} \cdot \sum\limits_{i \in [n]} 
 \thsindicbasevec{i,j}\cdot (\thsbasevec{i,j})
$$
\end{claim}

\begin{claim} \label{efs-revenue}
The revenue extracted by the fees $\vec{c_i}$ in $\mathcal{F}_S$ is at most -
$$
\sum\limits_{j=1}^m \sum\limits_{i \in \Asetforsubset{j}} \feemech{i,j}\cdot \frac{1-\delta}{\sizeActive} \cdot \chi_{i,j} + \sum\limits_{j\in [m]} \basevectorprob{j} \sum\limits_{i \in [n]} \thsindicbasevec{i,j} \cdot (v_{i,j} - \thsbasevec{i,j}).
$$
\end{claim}

Recall that the fees $\vec{c_i}$ are restricted by the profit that $\expostmechanism$ leaves player $i$ (Inequality~\eqref{chr_cond_matrix}) and thus we use $B$'s allocation to bound these fees.

\begin{proof}[Proof of Proposition~\ref{tigher-ub}]
Let $\expostmechanism, \vec{c_i}$ be the standard form of the given interim IR mechanism $M$ with agreement ratio of $x$ with $S$. 
By Claim~\ref{ex-post-revenue} and Claim~\ref{efs-revenue} we have:
\begin{equation*}
\begin{aligned}
     REV(M, \mathcal{F})  & \leq \sum\limits_{i=1}^{\sizeActive}{\thserd{i}}\cdot (\sum\limits_{\substack{j \in \Asetforplayer{i} \\ s.t. y_{i,j}\geq \thserd{i} }} {q_{i,\sigma_i(j)}\cdot \frac{1-\basevectorprob{}}{\sizeActive}})
     +  \sum\limits_{j \in [m]} \basevectorprob{j} \sum\limits_{i \in [n]} \thsindicbasevec{i,j}\cdot (\thsbasevec{i,j}) + \\
     &\sum\limits_{j=1}^m \sum\limits_{i \in \Asetforsubset{j}} \feemech{i,j}\cdot \frac{1-\delta}{\sizeActive} \cdot \chi_{i,j} + \sum\limits_{j\in [m]} \basevectorprob{j} \sum\limits_{i \in \Asetforsubset{j}} \thsindicbasevec{i,j} \cdot (v_{i,j} - \thsbasevec{i,j})
\end{aligned}
\end{equation*}
By definition, if $\thsindicbasevec{i,j} = 1$ then $v_{i,j} > \thsbasevec{i,j}$ and otherwise $\thsindicbasevec{i,j} = 0$ and recall that for every $j \in [m]$, only one player can have $\thsindicbasevec{i,j} =1$. Thus:
\begin{equation*}
    REV(M, \mathcal{F})   \leq\sum\limits_{i=1}^{\sizeActive}{\thserd{i}}\cdot (\sum\limits_{\substack{j \in \Asetforplayer{i} \\ s.t. y_{i,j}\geq \thserd{i} }} {q_{i,\sigma_i(j)}\cdot \frac{1-\basevectorprob{}}{\sizeActive}}) + \sum\limits_{j \in [m]} \basevectorprob{j} \cdot \mathop{\max}\limits_{i\in [n]}\{v_{i,j}\} + \sum\limits_{j=1}^m \sum\limits_{i \in \Asetforsubset{j}} \feemech{i,j}\cdot \frac{1-\delta}{\sizeActive} \cdot \chi_{i,j}
\end{equation*}
\end{proof}

The proof of Claim~\ref{ex-post-revenue} is simpler whereas the proof of Claim \ref{efs-revenue} is more involved (Section~\ref{efs-analysis}). 

\begin{proof} [Proof of Claim~\ref{ex-post-revenue}]

By construction, the instances in the subdistribution $P$ (see Section~\ref{P}) constitute $\sizeActive$ equal revenue distributions, one for each active player. In each distribution, the revenue the mechanism $\expostmechanism$ extracts from the respective active player $i$ is $\thserd{i}\cdot (\sum\limits_{\substack{j \in \Asetforplayer{i} \\ s.t. y_{i,j}\geq \thserd{i} }} {q_{i,\sigma_i(j)}\cdot \frac{1-\basevectorprob{}}{\sizeActive}}) $.  
Since all other players in the subdistribution have arbitrarily small values, much smaller than $y_{i,0}$, their contribution to the revenue is negligible.
Hence $\expostmechanism$'s revenue from instances in the support of $P$ is at most $\sum\limits_{i=1}^{\sizeActive}{\thserd{i}}\cdot (\sum\limits_{\substack{j \in \Asetforplayer{i} \\ s.t. y_{i,j}\geq \thserd{i} }} {q_{i,\sigma_i(j)}\cdot \frac{1-\basevectorprob{}}{\sizeActive}})  + \epsilonvalintermidate$.

All instances in the support of $O$ (Section~\ref{O}) and $R$ (Section~\ref{R}) have very small probability of $\epsilonprob{}$ and even if $\expostmechanism$ extracts all the social welfare from $O$ and $R$ as revenue is is still at most $\epsilonvalintermidate$.

Now, we bound the revenue from the instances in the subdistribution $E$ (Section~\ref{E}).
Consider some such instance $\Vec{v_j}$.
$\expostmechanism$ gives the item to player $i$ in $\vec{v_j}$ if player's $i$ value, $v_{i,j}$ is larger than the threshold that $\expostmechanism$ assigned player $i$ for $v_{-i} = \Vec{(v_j)}_{-i}$.
Now, since $\expostmechanism$ cannot give the item to more than one player in $\vec{v_j}$, at most one of the players $i \in [n]$ can have a threshold for $v_{-i} = \Vec{(v_j)}_{-i}$ that is smaller than his value in $\vec{v_j}$. I.e., at most one of the players $i \in [n]$ has $\thsbasevec{i,j} < v_{i,j}$ and only this player can have $\thsindicbasevec{i,j} = 1$ (by the definition of the variable $\thsindicbasevec{i,j}$). Recall that a player with value higher than his threshold wins the item and pays the threshold value:

\begin{equation}
    REV(B, \vec{v_j}) = \basevectorprob{j}\sum\limits_{i \in [n]} \thsindicbasevec{i,j}\cdot \thsbasevec{i,j}
\end{equation}

Summing the bounds on the revenue from the subdistributions $P, E, O$ and $R$ we have for every small enough $\epsilonvalintermidate > 0$:
$$
REV(\expostmechanism,F) \leq 2\epsilonvalintermidate +\sum\limits_{i=1}^{\sizeActive}{\thserd{i}}\cdot (\sum\limits_{\substack{j \in \Asetforplayer{i} \\ s.t. y_{i,j}\geq \thserd{i} }} {q_{i,\sigma_i(j)}\cdot \frac{1-\basevectorprob{}}{\sizeActive}}) +  \sum\limits_{j \in [m]} \basevectorprob{j} \sum\limits_{i \in [n]} \thsindicbasevec{i,j}\cdot \thsbasevec{i,j}
$$
\end{proof}

\subsubsection{Proof of Claim~\ref{efs-revenue}: Fees in the Distribution $\mathcal{F}_S$} \label{efs-analysis}

Recall that we fixed some interim IR mechanism in the standard form $(\expostmechanism, \vec{c_i})$ and our goal in this section is to bound the revenue extracted by the fees $\vec{c_i}$. 

We start with proving a lemma (Lemma~\ref{lemma:efs_depends_on_S}) that helps us to bound the fees charged from player $i$ when $v_{-i}= \Vec{(v_j)}_{-i}$ for some base vector $v_j$ (Lemma~\ref{lemma:fees_base_vector}).
Then, we bound the expected fees that can be charged from player $i$ for every one of his values (Lemma~\ref{lemma:fees_bound_per_val}). We then conclude the proof of Claim~\ref{efs-revenue}.

\begin{lemma}
\label{lemma:efs_depends_on_S}
For every $ j\in [m]$, $i \in \Asetforsubset{j}$, $(v_j)_{-i}$, if the mechanism $\expostmechanism$ does not allocate the item to bidder $i$ in the instance $(u_{i,j}, (v_j)_{-i})$, then $c_i((v_j)_{-i}) \leq 0$. 
\end{lemma}

We defer the proof of Lemma~\ref{lemma:efs_depends_on_S} to Appendix~\ref{proof-of-lemma-4.14}. For every player $i$ and subset $S_j$ that $i$ is active in $i \in \Asetforsubset{j}$, let:
\begin{equation*}
     \feesbasevec{i,j} =  [(\feemech{i,j} \cdot \frac{1-\basevectorprob{}}{\sizeActive \cdot \basevectorprob{j}}) + \thsindicbasevec{i,j}\cdot(v_{i,j} - \thsbasevec{i,j})]\cdot \basevectorprob{j}\cdot \chi_{i,j}. 
\end{equation*}

Now, the next lemma (\ref{lemma:fees_base_vector}) show that this value bound some of the fees taken from player $i$.

\begin{lemma} \label{lemma:fees_base_vector}
For every subset $S_j$ and every $ i \in \Asetforsubset{j}$, the fees $c_i((v_j)_{-i})$  satisfy: 
$$
c_i((v_j)_{-i}) \leq \frac{\feesbasevec{i,j}}{\basevectorprob{j}}.
$$
\end{lemma}

\begin{lemma}\label{lemma:fees_bound_per_val}
Fix a player $i\in [n]$ and a value in his support $w_i \in D_i$. It holds that:
\begin{align*}
\Pr\nolimits_{\mathcal{F}_i}(v_i = w_i)\cdot {[CP_i(\mathcal{F})]_{(w_i)}\cdot \Vec{c_i}} \leq
\begin{cases}
\epsilonvallowest + \basevectorprob{t}\cdot \thsindicbasevec{i,t}\cdot(v_{i,t}- \thsbasevec{i,t}) & \text{if $w_i = v_{i,t}$ for some $t \in [m] \and i \notin A_t$};\\
\epsilonvallowest & \text{if $w_i \neq v_{i,j}$ for every $j \in A_i$};\\
\epsilonvallowest + \feesbasevec{i,j}  & \text{if $w_i = v_{i,j}$ for some $j \in A_i$.}
\end{cases}
\end{align*}
\end{lemma}

The proofs of Lemma~\ref{lemma:fees_base_vector} and Lemma~\ref{lemma:fees_bound_per_val} are deferred to Appendices \ref{proof-of-4.15} and \ref{proof-of-4.16}, respectively. Given Lemma~\ref{lemma:fees_bound_per_val},  Claim~\ref{efs-revenue} follows almost immediately. 
\begin{proof}[Proof of Claim~\ref{efs-revenue}]

\begin{align*}
\sum\limits_{i=1}^{n}\sum\limits_{w_i \in D_i}{\Pr\nolimits_{\mathcal{F}_i}(v_i = w_i)\cdot [CP_i(\mathcal{F})]_{(w_i)}\cdot \Vec{c_i}} & \underbrace{\leq}_{Lemma~\ref{lemma:fees_bound_per_val}} \, \sum\limits_{i=1}^{n} ( |D_i|\cdot \epsilonvallowest + \sum\limits_{j \in \Asetforplayer{i}} \feesbasevec{i,j} + \sum\limits_{j \in \{ [m] \setminus \Asetforplayer{i}\}}  \basevectorprob{j} \cdot \thsindicbasevec{i,j} \cdot (v_{i,j} - \thsbasevec{i,j})) \\
& \leq \epsilonvalintermidate + \sum\limits_{j\in [m]} (\sum\limits_{i \in \Asetforsubset{j}} 
[(\feemech{i,j} \cdot \frac{1-\basevectorprob{}}{\sizeActive \cdot \basevectorprob{j}}) + \thsindicbasevec{i,j}\cdot(v_{i,j} ) - \thsbasevec{i,j})]\cdot \basevectorprob{j}\cdot \chi_{i,j} + \\
& \sum\limits_{i \in \{ [n] \setminus \Asetforsubset{j}\}}  \basevectorprob{j} \cdot \thsindicbasevec{i,j} \cdot (v_{i,j} - \thsbasevec{i,j})) \\
& \leq \epsilonvalintermidate+  \sum\limits_{j=1}^m \sum\limits_{i \in \Asetforsubset{j}} \feemech{i,j}\cdot \frac{1-\delta}{\sizeActive} \cdot \chi_{i,j} + \sum\limits_{j\in [m]} \sum\limits_{i \in [n]} \basevectorprob{j} \cdot \thsindicbasevec{i,j}(v_{i,j} - \thsbasevec{i,j})
\end{align*}
for every small enough $0 < \epsilonvalintermidate < 0.1$.

\end{proof}




\bibliography{revenue}
\bibliographystyle{alpha}

\appendix

\section{Appendix for Section \ref{app-claims}}\label{appendix-pf-no-interim-structure}

\subsection{Proof of Claim~\ref{no-interim-structure}} \label{pf-no-interim-structure}

\begin{proof}[Proof of Claim~\ref{no-interim-structure}]
Fix $k, n$, for every $0 < \varepsilon < 1$ and every $m \in \mathbb{N}$ we construct an $m-divisible$ set  w.r.t. $S_1, \dots, S_m$ of size $(n-k)\cdot m$ with parameters $\sizeActive = n-k, g_{\text{avg}} = \frac{1}{1-\varepsilon}, \alpha_{\text{avg}} = 1+ \varepsilon$ 
such that in all the instance in $S$ the item is not allocated to one of the $k$-highest player.
After obtaining such sets we can apply Theorem~\ref{construction_thm} and get that the each set is interim IR rigid with an $\frac{1}{m(1-\varepsilon)} + \frac{1+\varepsilon}{n-k}$-almost linear revenue disagreement function, which approaches $\frac{1}{n-k}$ when $m$ approaches infinity. 

Now, we describe the construction of those $m$-divisible sets. 
For every $m \in \mathbb{N}$ we construct an m-divisible set  w.r.t. $S_1, \dots S_m$ in the following way. The active players are the first $n-k$ players, for every subset $S_j$ its base vector is: $$(\Vec{v_j})_i = \begin{cases*}
        \varepsilon^{1-j} & \text{if $ i \leq n-k $}\\
        x_j & \text{otherwise ($ n-k < i \leq n$).}
        \end{cases*}$$
Where $x_j = \varepsilon^{-j}(1+\varepsilon) + (1+\varepsilon)$. Let $S_j = \{((x_j-\varepsilon, (v_j)_{-i}), i)\, |\, i\leq n-k\}$.  
        
Note that each base vector is strictly larger than the previous ones (coordinate-wise) and thus the sparsity requirement is satisfied (\ref{sparsity}). In addition every player $i \leq n-k$ has exactly one value $u_{i,j}$ s.t $(u_{i,j}, (v_{j})_{-i}) \in S_j$ and every player $i> n-k$ has no such value,  and thus it is indeed a base set (\ref{base_set}) and its size is $n-k$. Therefore $S = \bigcup\limits_{j \in [m]} S_j$ is an $m$-divisible set  w.r.t. $S_1, \dots S_m$ of size $m\cdot(n-k)$ and we only need to analyze its parameters.

For every active player $i \leq n-k$, we have $g_i = \frac{1}{1-\max\limits_{1 \leq j\leq k_i -1}\{\frac{y_{i,j}}{y_{i,{j+1}}}\}} = \frac{1}{1-\varepsilon} $, then $g_{\text{avg}} = \frac{1}{1-\varepsilon}$.
For every subset $S_j$, we have $\alpha_j= \frac{\max\limits_{i \in [n]}\{v_{i,j}\}}{\min\limits_{k \in A_j}\{u_{k,j} -v_{k,j}\}}=\frac{x_j}{x_j-\varepsilon - \varepsilon^{1-j}} = 1+\varepsilon$, then $\alpha_{\text{avg}} = 1+\varepsilon$.

\end{proof}

\section{Missing Proofs from Section~\ref{proof_construction}}

We use the following notation; For every subset $S_j$ and every active player in it $i \in \Asetforsubset{j}$ let:

\begin{subequations}\label{not-cond-active}
    \begin{equation}
        \basevectorprob{i,j} := \Pr_{\mathcal{F}} ({v_{-i} = (v_j)_{-i} \, |\, v_i = v_{i,j}})
    \end{equation}
    \begin{equation}
        \epsilonprob{i,j} := \sum_{v_{-i} \neq (v_j)_{-i}, (\epsilonvaleqr{i}\dots \epsilonvaleqr{i})}\Pr_{\mathcal{F}} ({v_{-i} = (v_j)_{-i} \, |\, v_i = v_{i,j}})
    \end{equation}
\end{subequations}

\subsection{Proof of Lemma~\ref{lemma:y_0_comp}}\label{proof-of-4.2}




\begin{proof}[Proof of Lemma~\ref{lemma:y_0_comp}]

Consider the set of equations in~\eqref{minimal_efs_equal}. A positive solution to these equations exists: for example set $y_{1,0}=1$ and for every $1 < k\leq \sizeActive$
set $y_{k,0}=\frac{d_1}{d_k}$ which is also positive because $d_i > 0$ for every $i \leq \sizeActive$. Observe that multiplying a solution to these equations by some constant (i.e., multiplying each $y_{i,0}$ by the same constant) yields a valid solution as well. Thus, we can choose a solution that is positive and the maximal value in it is arbitrarily small.

We start with a positive solution $y_{1,0}, \dots y_{\sizeActive,0}$ to the set of equations in~\eqref{minimal_efs_equal}. For every active player $i$ for which if $y_{i,0} \geq y_{i,1}$, we decrease the value of the solution by multiplying it all by some positive constant such that the inequality will hold (for example, by $\frac{y_{i,1}}{2\cdot y_{i,0}}$). Now, we have a positive solution to~\eqref{minimal_efs_equal} that satisfies the inequalities in~\eqref{y1>y0}. We also need to also satisfy the inequalities in~\eqref{efs_constraint}. 
Observe that if $y_{1,0}\leq \frac{\frac{1}{m}(\basevectorprob{}-\epsilonvallowest)\sizeActive}{(1-\basevectorprob{})d_1} \cdot \min\limits_{j\in [m] \atop i \in A_j}\{\thsintermediate{i,j}-v_{i,j}\}$ then the inequalities in~\eqref{efs_constraint} are satisfied. Note that the right hand side is a positive number. Thus, if $y_{1,0}$ is not small enough we can multiply the solution by a small positive constant such that the new $y_{1,o}$ will be small enough. Therefore, there exists a positive solution to~\eqref{minimal_efs_equal} that satisfies the inequalities in~\eqref{delta_constraint}. 
\end{proof} 

\subsection{Proof of Lemma~\ref{lemma:EQD_probabilites}}\label{proof-of-4.3}



\begin{proof}[Proof of Lemma~\ref{lemma:EQD_probabilites}]
By applying simple algebraic manipulations on the equations in~\eqref{EQD}, we get that they are equivalent to the following set of equations:
\begin{align}
\begin{split}
    & q_{i,j} =  \frac{y_{i,0}}{y_{i,j}}-\frac{y_{i,0}}{y_{i,j+1}} \quad \forall \, 1 \leq j \leq k_1 \\
    & q_{i,k_i+1}  = \frac{y_{i,0}}{y_{i,k_i+1}} \\
    & q_{i,0}  =  1 - (q_{i,1} + \dots q_{i,k_i+1})
\end{split}
\end{align}

Observe that for every $1 \leq j \leq k_i+1 $ it holds that $0 < q_{i,j} < 1$ since $0 < y_{i,0} < y_{i,1} < \dots < y_{i,k_i+1}$. From Equation~\ref{EQD} for $j=1$ we also have: 
$$
0 < q_{i,1} + \dots + q_{i,k_i+1} = \frac{y_{i,0}}{y_{i,1}} < 1
$$
Then, also $q_{i,0} = 1 - (q_{i,1} + \dots q_{i,k_i+1})$ satisfies $0 < q_{i,0} < 1$ and by definition $q_{i,0} + q_{i,1} + \dots + q_{i,k_i+1} = 1$. 
\end{proof}

\subsection{Proof of Lemma~\ref{lemma:expected_sw_bound}}\label{proof-of-4.4}


\begin{proof}[Proof of Lemma~\ref{lemma:expected_sw_bound}]
 From Equation~\ref{EQD} we have:
 \begin{align*}
    y_{i,0} = y_{i,j+1} \cdot (q_{i,{j+1}} + \dots q_{i,{k_i+1}}) & \iff q_{i,{j+1}} + \dots q_{i,k_i+1} = \frac{y_{i,0}}{y_{i,j+1}} \\
    y_{i,0} = y_{i,j} \cdot (q_{i,j} + \dots q_{i,k_i+1}) & = y_{i,j} \cdot q_{i,j} + y_{i,j}\cdot \frac{y_{i,0}}{y_{i,j+1}}\\
    y_{i,j} \cdot q_{i,j} = y_{i,0} \cdot (1-& \frac{y_{i,j}}{y_{i,j+1}}) \geq y_{i,0} \cdot d_i
 \end{align*}
\end{proof}


\subsection{Proof of Lemma~\ref{lemma:legal-interim}}\label{proof-of-4.11}

In this section we use the notation defined in~\eqref{not-cond-active}.




\begin{proof}
By Claim~\ref{legal-ex-post}, $\expostmechanism'$ is a deterministic, dominant strategy incentive compatible and ex-post IR mechanism. By Proposition~\ref{characterization} it suffices to show that:
$${[CP_i(\mathcal{F})\cdot \vec{c'_i}]}_k  \leq \pi_i^{{\expostmechanism'(\mathcal{F})}}(v_{i,k}) \quad \quad  \forall i \in [n], \, \forall k=1, \dots, |\mathcal{F}_i|$$

Since the only non-zero fees are $c'_i((v_j)_{-i})$ for an active player $i \in A$ and a subset $S_j$ that he is active in, we only need to consider values $v_i \in D_i$ for which these values have non-zero probability, i.e.,  $[CP_i(\mathcal{F})\cdot \vec{c'_i}]_{(v_i,(v_j)_{-i})} > 0$. These values are $v_{i,j}, {u'}_{i,j}$ and $u_{i,j}$.

\begin{enumerate}
    \item \label{v-lb} $[CP_i(\mathcal{F})\cdot c'_i]_{(v_{i,j})}  \leq \pi_i^{{\expostmechanism'(\mathcal{F})}}(v_{i,j})$.
    \item \label{u'-lb} $[CP_i(\mathcal{F})\cdot c'_i]_{({u'}_{i,j})} \leq \pi_i^{{\expostmechanism'(\mathcal{F})}}({u'}_{i,j})$.
    \item \label{u_lb} $[CP_i(\mathcal{F})\cdot c'_i]_{({u}_{i,j})} \leq \pi_i^{{\expostmechanism'(\mathcal{F})}}({u}_{i,j})$.
\end{enumerate}

\begin{enumerate}
    \item \begin{align*}
         [CP_i(\mathcal{F})\cdot c'_i]_{(v_{i,j})} = & \, \basevectorprob{i,j}\cdot c'_i((v_j)_{-i}) = \basevectorprob{i,j}\cdot \frac{\frac{1-\basevectorprob{}}{\sizeActive}\cdot e}{\basevectorprob{j}} \underbrace{\leq}_{Lemma~\ref{lemma:expected_sw_bound}} \basevectorprob{i,j}\cdot \frac{\frac{1-\basevectorprob{}}{\sizeActive}\cdot (q_{i,\sigma_i(j)}\cdot y_{i,\sigma_i(j)})}{\basevectorprob{j}} \\
         =& \, \frac{\frac{(1-\basevectorprob{})}{\sizeActive}(q_{i,\sigma_i(j)}\cdot y_{i,\sigma_i(j)})}{\Pr\nolimits_{\mathcal{F}_i}(v_i=v_{i,j})}
         = y_{i,\sigma_i(j)} -  \frac{y_{i,\sigma_i(j)}\cdot(\Pr\nolimits_{\mathcal{F}_i}(v_i=v_{i,j})-\frac{(1-\basevectorprob{})}{\sizeActive}(q_{i,\sigma_i(j)}\cdot y_{i,\sigma_i(j)}))}{\Pr\nolimits_{\mathcal{F}_i}(v_i=v_{i,j})} \\
        = & \,  y_{i,\sigma_i(j)} - \frac{ y_{i,\sigma_i(j)}(\basevectorprob{i,j} + \epsilonprob{i,j})\cdot (\Pr\nolimits_{\mathcal{F}_i}(v_i=v_{i,j}))}{\Pr\nolimits_{\mathcal{F}_i}(v_i=v_{i,j})} \\
        = & y_{i,\sigma_i(j)} \cdot (1- \basevectorprob{i,j} - \epsilonprob{i,j}) \leq \pi_i^{{\expostmechanism'(\mathcal{F})}}(v_{i,j})
    \end{align*}
    \item \label{u'} \begin{align*}
        [CP_i(\mathcal{F})\cdot c'_i]_{({u'}_{i,j})} = & c'_i((v_j)_{-i}) =  \cfrac{\frac{1-\basevectorprob{}}{\sizeActive}\cdot e}{\basevectorprob{j}} =  \cfrac{1-\basevectorprob{}}{\sizeActive}\cdot e : \cfrac{(1-\basevectorprob{})\cdot e}{\sizeActive\cdot \min\limits_{k \in \Asetforsubset{j}}{\{{u'}_{k,j}-v_{k,j} \}}} \\
        = & \min\limits_{k \in \Asetforsubset{j}}{\{{u'}_{k,j}-v_{k,j}\}} \leq {u'}_{i,j}-v_{i,j} \leq \pi_i^{{\expostmechanism'(\mathcal{F})}}({u'}_{i,j})
    \end{align*}
    \item \begin{align*}
        [CP_i(\mathcal{F})\cdot c'_i]_{({u}_{i,j})} = & c'_i((v_j)_{-i}) \underbrace{\leq}_{\text{Item } \ref{u'}} {u'}_{i,j}-v_{i,j} \leq  {u}_{i,j}-v_{i,j} \leq \pi_i^{{\expostmechanism'(\mathcal{F})}}({u}_{i,j})
    \end{align*}
\end{enumerate}

\end{proof}

\subsection{Proof of Lemma~\ref{lemma:efs_depends_on_S}}\label{proof-of-lemma-4.14}


\begin{proof}[Proof of Lemma ~\ref{lemma:efs_depends_on_S}]
By the uniqueness of thresholds property of $S$ (Definition~\ref{uniqueness}), $u_{i,j}$ is different than any other value in the support of player $i$. Then, according to the construction, the row vector $[CP_i(\mathcal{F})]_(u_{i,j})$ in the conditional probability matrix of player that corresponds to the value $u_{i,j}$ has 0 entries everywhere except for the column that corresponds to the valuation $\vec{v_j}_{-i}$ of the other players. 
Therefore:

\begin{subequations} 
\begin{equation}\label{u_value_prof}
    \pi_i^{\expostmechanism(\mathcal{F})}({{u}_{i,j}}) = \pi_i^{\expostmechanism(\mathcal{F})}({{u}_{i,j}}, \vec{v_j}_{-i})
\end{equation}
\begin{equation}\label{u_value_fee}
    {[CP_i(\mathcal{F})]_{(u_{i,j})}\cdot \Vec{c_i}} = c_i(\vec{v_j}_{-i})
\end{equation}
\end{subequations}

Assume that $\expostmechanism$ does not allocate the item to player $i\in \Asetforsubset{j}$ in the instance $(u_{i,j}, (v_j)_{-i})$. Then
\begin{equation} \label{profit_0_u}
    \pi_i^{\expostmechanism(\mathcal{F})}({{u}_{i,j}}, \vec{v_j}_{-i}) = 0
\end{equation}

By Equation~\ref{characterizarion_condition}  we have:
\begin{equation} \label{fees_by_profit_u}
{[CP_i(\mathcal{F})]_{(u_{i,j})}\cdot \Vec{c_i}} \leq \pi_i^{\expostmechanism(\mathcal{F})}({{u}_{i,j}})
\end{equation}

Combining Equation~\ref{u_value_prof}, Equation~\ref{u_value_fee},  Equation~\ref{profit_0_u}, and Equation~\ref{fees_by_profit_u}, we conclude that:

$$
c_i(\vec{v_j}_{-i}) \leq \pi_i^{\expostmechanism(\mathcal{F})}({{u}_{i,j}}, \vec{v_j}_{-i}) = 0
$$

\end{proof}

\subsection{Proof of Lemma~\ref{lemma:fees_base_vector}}\label{proof-of-4.15}

In this section we use the notation defined in~\eqref{not-cond-active}.

\begin{proof}[Proof of Lemma~\ref{lemma:fees_base_vector}]
Observe that when $v_i= {u'}_{i,j}$ it holds that $v_{-i} = (v_j)_{-i}$ with probability 1. Therefore:
\begin{subequations}
\begin{equation} \label{fees_base_by_u'}
    c_i((v_j)_{-i}) \leq \pi^{\expostmechanism(\mathcal{F})}({u'}_{i,j})
\end{equation}
\begin{equation} \label{u'_profit}
    \pi^{\expostmechanism(\mathcal{F})}({u'}_{i,j}) = \pi^{\expostmechanism(\mathcal{F})}({u'}_{i,j}, \Vec{v_j}_{-i}) \leq  u'_{i,j} - v_{i,j} + \thsindicbasevec{i,j}\cdot(v_{i,j} - \thsbasevec{i,j})
\end{equation}
\end{subequations}

Now, by the definition of $u'_{i,j}$ in Section~\ref{O} and the definition of $\basevectorprob{j}$ in Section~\ref{E}, we have:
\begin{equation} \label{u'_pro_bound}
     u'_{i,j} - v_{i,j} \leq v_{i,j} + \min\limits_{k \in \Asetforsubset{j}} \{\thsintermediate{k,j}- v_{k,j}\} -v_{i,j} = \frac{(1-\basevectorprob{})\cdot e}{\sizeActive \cdot \basevectorprob{j}}
\end{equation}

Combining Equation~\ref{fees_base_by_u'}, Equation~\ref{u'_profit}, and Equation~\ref{u'_pro_bound} we get:
\begin{equation*}
    c_i((v_j)_{-i}) \leq \frac{(1-\basevectorprob{})\cdot e}{\sizeActive \cdot \basevectorprob{j}} + \thsindicbasevec{i,j}\cdot(v_{i,j} - \thsbasevec{i,j})
\end{equation*}

Now, by Lemma~\ref{lemma:efs_depends_on_S}, if $\expostmechanism$ does not allocate player $i$ the item in the instance $(u_{i,j}, \Vec{v_j}_{-i})$ (i.e., $\chi_{i,j} = 0$) then $ c_i((v_j)_{-i}) \leq 0$, and:

\begin{equation}\label{e_ub}
    c_i((v_j)_{-i}) \leq \chi_{i,j} \cdot [\frac{(1-\basevectorprob{})\cdot e}{\sizeActive \cdot \basevectorprob{j}} + \thsindicbasevec{i,j}\cdot(v_{i,j} - \thsbasevec{i,j})] 
\end{equation}

We finish by proving that:
\begin{equation}
     c_i((v_j)_{-i}) \leq \chi_{i,j} \cdot [\frac{(1-\basevectorprob{})\cdot (y_{i,\sigma_i(j)}-\thserd{i})\cdot q_{i,\sigma_i(j)}}{\sizeActive \cdot \basevectorprob{j}} + \thsindicbasevec{i,j}\cdot(v_{i,j} - \thsbasevec{i,j})] 
\end{equation}

By the characterization~\ref{characterization} we have:

\begin{equation*}
    [CP_i(\mathcal{F})\cdot c_i]_{(v_{i,j})}  \leq \pi_i^{{\expostmechanism(\mathcal{F})}}(v_{i,j})
\end{equation*}

Then:

\begin{equation*}
     \basevectorprob{i,j}\cdot c_i((v_j)_{-i}) \leq [CP_i(\mathcal{F})\cdot c_i]_{(v_{i,j})} \leq \pi_i^{{\expostmechanism(\mathcal{F})}}(v_{i,j}) \leq \frac{q_{i, \sigma_i(j)}\cdot\frac{(1-\basevectorprob{})}{\sizeActive}\cdot (y_{i,\sigma_i(j)}-\thserd{i})}{\Pr\nolimits_{\mathcal{F}_i}(v_i=v_{i,j})} + (\thsindicbasevec{i,j}\cdot(v_{i,j} - \thsbasevec{i,j}))\basevectorprob{i,j} +v_{i,j}\cdot\epsilonprob{i,j}
\end{equation*}

For every small enough $\epsilonvallowest>0$ we get:
\begin{equation*}
     c_i((v_j)_{-i}) \leq [\frac{(1-\basevectorprob{})\cdot (y_{i,\sigma_i(j)}-\thserd{i})\cdot q_{i,\sigma_i(j)}}{\sizeActive \cdot \basevectorprob{j}} + \thsindicbasevec{i,j}\cdot(v_{i,j} - \thsbasevec{i,j})] +\epsilonvallowest
\end{equation*}

Lastly, by Lemma~\ref{lemma:efs_depends_on_S}, if $\expostmechanism$ does not allocate player $i$ the item in the instance $(u_{i,j}, \Vec{v_j}_{-i})$ (i.e., $\chi_{i,j} = 0$) then $ c_i((v_j)_{-i}) \leq 0$, and:

\begin{equation}
     c_i((v_j)_{-i}) \leq \chi_{i,j} \cdot [\frac{(1-\basevectorprob{})\cdot (y_{i,\sigma_i(j)}-\thserd{i})\cdot q_{i,\sigma_i(j)}}{\sizeActive \cdot \basevectorprob{j}} + \thsindicbasevec{i,j}\cdot(v_{i,j} - \thsbasevec{i,j})] 
\end{equation}

\end{proof}

\subsection{Proof of Lemma~\ref{lemma:fees_bound_per_val}}\label{proof-of-4.16}

To prove Lemma~\ref{lemma:fees_bound_per_val}, we need the following observation. 

\begin{observation}\label{bound_on_every_v_-i}
Let $w_i = v_{i,j}$ for some $j \in [m]$ that player $i$ is active in.
Consider some value $v_{-i} = w_{-i}$ that has non-zero probability when $v_i = w_i$, (i.e., $\Pr\nolimits_{F}(v_{-i} = w_{-i} \, |\,  v_i = w_i) > 0 $), and that satisfies $w_{-i} \neq \Vec{v_k}_{-i}$ for every subset $S_k$ that player $i$ is active in. Then, by construction (specifically, the subdistribution R that is defined in Section~\ref{R}), this value of $v_{-i}$ is paired with some small value of player $i$,  $z_i \leq \epsilonvallowest$ such that $\Pr\nolimits_{F}(v_{-i} = w_{-i} \, |\,  v_i = z_i) = 1 $ (see Section~\ref{R})


\end{observation}


\begin{proof}[Proof of Lemma~\ref{lemma:fees_bound_per_val}]
We prove the lemma for each possible value $w_i \in D_i$. For every value $w_i$, we consider the row in the conditional probability matrix of player $i$ that corresponds to this value, $[CP_i(\mathcal{F})]_{(w_i)}$, and use Observations~\ref{efs_bounded_by_value} and \ref{efs_upper_bound} to prove the required bound.

\begin{itemize}
    \item  \emph{$w_i = v_{i,t}$  for some subset $S_t$ that player $i$ is not active in}. This value satisfies the first condition in the statement of the lemma. 
    By the uniqueness of thresholds property (Definition~\ref{uniqueness}), the row in the conditional probability matrix of player $i$ that corresponds to the value $v_{i,t}$,
    $[CP_i(\mathcal{F})]_{(v_{i,t})}$ can have non-zero entries only in the columns that corresponds to  $v_{-i} \neq \vec{v_j}_{-i}$ for every subset $S_j$ that player $i$ is active in. 
    We bound the expected fees taken from player $i$ when his value is $v_{i,t}$ by the his expected profit in the mechanism $\expostmechanism$ when his value is $v_{i,t}$ (See~\ref{chr_cond_matrix}). 
    $$
    \Pr\nolimits_{\mathcal{F}_i}(v_i = v_{i,t})\cdot {[CP_i(\mathcal{F})]_{(v_{i,t})}\cdot \Vec{c_i}}  \leq 
    \epsilonvallowest + \basevectorprob{j}\cdot\thsindicbasevec{i,j}\cdot(v_{i,j} - \thsbasevec{i,j}).
    $$
    \item \emph{$w_i = y_{i,0}$ or $w_i = y_{i,k+1}$}. If $w_i \leq \epsilonvallowest$, then due to Observation~\ref{efs_bounded_by_value} we get the required bound. Otherwise, by Observation~\ref{bound_on_every_v_-i} (as $w_i > \epsilonvallowest$) and Observation~\ref{efs_upper_bound} we get:
    $$
    \Pr\nolimits_{\mathcal{F}_i}(v_i = v_{i,t})\cdot {[CP_i(\mathcal{F})]_{(v_{i,t})}\cdot \Vec{c_i}}  \leq \Pr\nolimits_{\mathcal{F}_i}(v_i = v_{i,t}) \cdot \epsilonvallowest \leq \epsilonvallowest.
    $$
    \item \emph{$w_i \neq v_{i,k}$ for every subset $S_k$ and $w_i \neq  y_{i,0},  y_{i,k_i+1}.$} Then, $\Pr\nolimits_{\mathcal{F}_i}(v_i = w_i) = \epsilonprob{}$ and since $\epsilonprob{} \cdot \max\limits_{j \in [m]} \{ \max\limits_{i \in \Asetforsubset{j}} \{u_{i,j}\} \} \leq \epsilonvallowest$ (by the condition in Equation~\ref{smaller_than_lowest}) we get:
    $$
    \Pr\nolimits_{\mathcal{F}_i}(v_i = w_i)\cdot {[CP_i(\mathcal{F})]_{(w_i)}\cdot \Vec{c_i}} \underbrace{\leq}_{Observation~\ref{efs_bounded_by_value}}\Pr\nolimits_{\mathcal{F}_i}(v_i = w_i)\cdot w_i \leq \epsilonvallowest.
    $$
    \item \emph{$w_i = v_{i,j}$ for some subset $S_j$ that player $i$ is active in.}
    By the uniqueness of thresholds property (Definition~\ref{uniqueness}), the row in the conditional probability matrix of player $i$ that corresponds to the value $v_{i,j}$,
    $[CP_i(\mathcal{F})]_{(v_{i,j})}$ can have non-zero entries only in the columns that corresponds to  $v_{-i} \neq \vec{v_k}_{-i}$ for every subset $S_k$, different from $S_j$ and that player $i$ is active in (i.e., $k\neq j$ and $k \in A_i$). Now, we can apply Observation~\ref{bound_on_every_v_-i} (as $v_{i,j} > \epsilonvallowest$) and Observation~\ref{efs_upper_bound} and get:
    \begin{align*}
    \Pr\nolimits_{\mathcal{F}_i}(v_i = v_{i,j})\cdot {[CP_i(\mathcal{F})]_{(v_{i,j})}\cdot \Vec{c_i}} & \leq  \Pr\nolimits_{\mathcal{F}_i}(v_i = v_{i,j
    })\cdot 
    (\basevectorprob{i,j}\cdot [\vec{c_i}]_{({v_j}_{-i})} + \epsilonvallowest) \\
    & = \basevectorprob{j} \cdot [\vec{c_i}]_{({v_j}_{-i})} + \epsilonvallowest
    \underbrace{\leq}_{Lemma~\ref{lemma:fees_base_vector}} \epsilonvallowest + \feesbasevec{i,j} 
    \end{align*}
\end{itemize}
Since we assumed uniqueness of thresholds (Definition~\ref{uniqueness}), these cases cover all possible cases.  
\end{proof}
\section{Almost Linear}\label{al-sec}

We show that for values of $x \in [1-\contconstruction,1)$ the approximation ratio of a mechanism with agreement ratio at most $x$ is less than 1 (Claim~\ref{almost-linear-claim}). For that purpose we will need a tighter analysis of the revenue of the optimal mechanism (Proposition~\ref{lb-rev-almost-linear}).

We use the following two propositions to derive Claim~\ref{almost-linear-claim}.

\begin{proposition}\label{lb-rev-almost-linear}
There exists a deterministic, dominant strategy incentive compatible, and interim IR mechanism which extracts in expectation over $\mathcal F_S$ revenue of at least: 

$$ 
\sum\limits_{i=1}^{\sizeActive}{y_{i,0} \cdot \frac{1-\basevectorprob{}}{\sizeActive}} + \sum\limits_{j=1}^{m} \basevectorprob{j}\cdot \max\limits_{i \in {n}}\{v_{i,j}\} +
\sum\limits_{j=1}^m \sum\limits_{i \in \Asetforsubset{j}} \feeopt{i,j}\cdot \frac{1-\delta}{\sizeActive}
$$ 

Where $\feeopt{i,j} = \min \{ (y_{i,\sigma_i(j)}-y_{i,0})\cdot q_{i,\sigma_i(j)}, \, e\}$ for every active player $i \in A$ and every $j \in \Asetforplayer{i}$.
\end{proposition}

\begin{proposition}\label{ub-rev-almost-linear}
Let $M$ be a deterministic, dominant strategy incentive compatible, and interim IR mechanism with agreement ratio of $x$ with $S$. The expected revenue of $M$ over $\mathcal{F}_S$ is at most:
$$
\sum\limits_{i=1}^{\sizeActive}{\thserd{i}}\cdot (\sum\limits_{\substack{j \in \Asetforplayer{i} \\ s.t. y_{i,j}\geq \thserd{i} }} {q_{i,\sigma_i(j)}\cdot \frac{1-\basevectorprob{}}{\sizeActive}}) + \sum\limits_{j=1}^{m} \basevectorprob{j}\cdot \max\limits_{i \in {n}}\{v_{i,j}\} + \sum\limits_{j=1}^m \sum\limits_{i \in \Asetforsubset{j}} \feemech{i,j}\cdot \frac{1-\delta}{\sizeActive} \cdot \chi_{i,j}
$$

where $\thserd{i}$ is the threshold of player $i$ for $v_{-i} = (\epsilonvalrestrictefs{i},\dots,\epsilonvalrestrictefs{i})$, $\feemech{i,j} = \min \{ (y_{i,\sigma_i(j)}-\thserd{i})\cdot q_{i,\sigma_i(j)}, \, e\}$ for every active player $i \in A$ and every $j \in \Asetforplayer{i}$, and $\chi_{i,j}$ is an indicator variable that equals $1$ if $M$ agrees with $S$ on the allocation in the instance $({u}_{i,j}, (v_j)_{-i})$ and $0$ otherwise.
\end{proposition}

\begin{proof}[Proof of Claim~\ref{almost-linear-claim}]
By Proposition~\ref{lb-rev-almost-linear} and Proposition~\ref{ub-rev-almost-linear}, it suffices to show that:
$$
\sum\limits_{i=1}^{\sizeActive}{y_{i,0} \cdot \frac{1-\basevectorprob{}}{\sizeActive}} + 
\sum\limits_{j=1}^m \sum\limits_{i \in \Asetforsubset{j}} \feeopt{i,j}\cdot \frac{1-\delta}{\sizeActive} \underbrace{>}_{for \, x< 1}
\sum\limits_{i=1}^{\sizeActive}{\thserd{i}}\cdot (\sum\limits_{\substack{j \in \Asetforplayer{i} \\ s.t. y_{i,j}\geq \thserd{i} }} {q_{i,\sigma_i(j)}\cdot \frac{1-\basevectorprob{}}{\sizeActive}})+ \sum\limits_{j=1}^m \sum\limits_{i \in \Asetforsubset{j}} \feemech{i,j}\cdot \frac{1-\delta}{\sizeActive} \cdot \chi_{i,j}
$$

We show that for every $i \in A$ we have (Lemma~\ref{tech-lemma-almost-linear}) : 
$$
y_{i,0} \cdot \frac{1-\basevectorprob{}}{\sizeActive} + \sum\limits_{j\in \Asetforplayer{i}} \feeopt{i,j}\cdot \frac{1-\delta}{\sizeActive} \underbrace{\geq}_{for \, x< 1}
\thserd{i}\cdot (\sum\limits_{\substack{j \in \Asetforplayer{i} \\ s.t. y_{i,j}\geq \thserd{i} }} {q_{i,\sigma_i(j)}\cdot \frac{1-\basevectorprob{}}{\sizeActive}})+ \sum\limits_{j\in \Asetforplayer{i}} \feemech{i,j}\cdot \frac{1-\delta}{\sizeActive}.
$$

Then, since $x< 1$, at least one of the $\chi_{i,j}$ is equal 0 and we get a strict inequality and the claim follows. 

\begin{lemma}\label{tech-lemma-almost-linear}
for every $i \in A$ it holds that:
$$
y_{i,0} \cdot \frac{1-\basevectorprob{}}{\sizeActive} + \sum\limits_{j\in \Asetforplayer{i}} \feeopt{i,j}\cdot \frac{1-\delta}{\sizeActive} \underbrace{\geq}_{for \, x< 1}
\thserd{i}\cdot (\sum\limits_{\substack{j \in \Asetforplayer{i} \\ s.t. y_{i,j}\geq \thserd{i} }} {q_{i,\sigma_i(j)}\cdot \frac{1-\basevectorprob{}}{\sizeActive}})+ \sum\limits_{j\in \Asetforplayer{i}} \feemech{i,j}\cdot \frac{1-\delta}{\sizeActive}.
$$
\end{lemma}

\begin{proof}[Proof of Lemma~\ref{tech-lemma-almost-linear}]
Observe that $\thserd{i}\cdot (\sum\limits_{\substack{j \in \Asetforplayer{i} \\ s.t. y_{i,j}\geq \thserd{i} }} {q_{i,\sigma_i(j)}\cdot \frac{1-\basevectorprob{}}{\sizeActive}})$ is the ex-post revenue mechanism $M$ extracts from the equal revenue distribution of player $i$ and thus it holds that :
$$\thserd{i}\cdot (\sum\limits_{\substack{j \in \Asetforplayer{i} \\ s.t. y_{i,j}\geq \thserd{i} }} {q_{i,\sigma_i(j)}\cdot \frac{1-\basevectorprob{}}{\sizeActive}}) \leq y_{i,0} \cdot \frac{1-\basevectorprob{}}{\sizeActive}.$$
If $t_i \geq y_{i,0}$, then by definition $\feeopt{i,j} \geq \feemech{i,j}$ and the claim holds. 
Now, we assume that $t_i < y_{i,0}$ and prove that for every $j \in \Asetforplayer{i}$ it holds that 
\begin{equation}\label{tech-lemma-al-for-j}
    q_{i,\sigma_i(j)}\cdot y_{i,0}+ \feeopt{i,j} \geq q_{i,\sigma_i(j)}\cdot \thserd{i} + \feemech{i,j}
\end{equation}
which will conclude the proof of the lemma.

If $\feemech{i,j}=(y_{i,\sigma_i(j)}-t_i)\cdot q_{i,\sigma_i(j)}$, then $\feeopt{i,j} = (y_{i,\sigma_i(j)}-y_{i,0})\cdot q_{i,\sigma_i(j)}$ and Inequality ~\eqref{tech-lemma-al-for-j} holds.
Otherwise $\feemech{i,j}= e$, now if also $\feeopt{i,j} = e$ then Inequality  ~\eqref{tech-lemma-al-for-j} holds since we assumed that $\thserd{i} < y_{i,0}$. 
We are left with the case that $\feemech{i,j}= e$ and  $\feeopt{i,j} = (y_{i,\sigma_i(j)}-y_{i,0})\cdot q_{i,\sigma_i(j)}$, then:
\begin{align*}
    q_{i,\sigma_i(j)}\cdot \thserd{i} + \feemech{i,j} & \leq q_{i,\sigma_i(j)}\cdot \thserd{i} +  (y_{i,\sigma_i(j)}-\thserd{i})\cdot q_{i,\sigma_i(j)} = y_{i,\sigma_i(j)}\cdot q_{i,\sigma_i(j)}\\
    & =  q_{i,\sigma_i(j)}\cdot y_{i,0} + (y_{i,\sigma_i(j)}-y_{i,0})\cdot q_{i,\sigma_i(j)} =  q_{i,\sigma_i(j)}\cdot y_{i,0} +\feeopt{i,j}\\
\end{align*}

and Inequality~\eqref{tech-lemma-al-for-j} holds.
\end{proof}

\end{proof}

Observe that Proposition~\ref{ub-rev-almost-linear} is proved in Section~\ref{upper_bound_sec}. The proof of Proposition~\ref{lb-rev-almost-linear} is provided in the next section. 

\subsection{Proof of Proposition~\ref{lb-rev-almost-linear}}

We prove this proposition by providing an ex-post IR, deterministic and dominant strategy incentive compatible mechanism $\expostmechanism'$ with fees $\vec{c'_i}$ that together compose an interim IR mechanism $M'$ with the required revenue.   

We start with describing the allocation function of $\expostmechanism'$ and its payments in the instances of $\mathcal F_S$ by specifying its thresholds. Recall that in an ex-post IR mechanism the payment of a winning bidder equals his threshold and that a losing bidder pays $0$. 

For every active player $i \in A$, let player $i$'s threshold for $v_{-i} = (\epsilonvalrestrictefs{i},\dots,\epsilonvalrestrictefs{i})$ be $y_{i,0}$.
All other thresholds are set to $\infty$, i.e., $\expostmechanism'$ does not allocate the item in any other case. 

For every instance $v_j$ (for $j \in [m]$), let the highest player have a threshold equal his value in $v_j$ (break tie arbitrarily). For every other active player $i\in A$, set his threshold for $v_{-i}=(v_j)_{-i}$ to be $v_{i,j} + \epsilon$ for some arbitrarily small value of $\epsilon > 0$.

All other thresholds are set to $\infty$, i.e., $\expostmechanism'$ does not allocate the item in any other case. 

\begin{claim} \label{legal-ex-post-lb}
$\expostmechanism'$ is a deterministic, ex-post IR , and dominant strategy incentive compatible mechanism.
\end{claim}

\begin{proof}[Proof of Claim~\ref{legal-ex-post-lb}]
$\expostmechanism'$ is clearly a deterministic, ex-post IR mechanism. It is also a dominant strategy mechanism since each player is allocated the item if his value is more than some threshold that does not depend on his value. It remains to show that the mechanism is feasible, i.e., the item is not allocated to two players in the same instance.

Consider some instance $\vec{v}$. If $\expostmechanism'$ allocates the item to some player $i$ in $\vec v$, then when the values of the other players are $\vec{v}_{-i}$, the threshold of player $i$ is some $t_i \leq \vec{v}_i$.

Observe that for every instance $\vec{v}$ in the support of the subdistribution $P$,  $\expostmechanism'$ only 
allocates the item to the $i$'th player. For every other player $j \neq i$ the values of $v_{-j}$ of the instances in this subdistribution are unique. Thus, the threshold of every $j \neq i$ is $\infty$. 

The other case we need to consider is that $v_{-1} = (\vec{v_j})_{-1}$ and $v_{-2} = (\vec{v_k})_{-2}$ for some $j,k \in [m]$. However, since the base vectors $\vec{v_1}, \dots, \vec{v_m}$ are sparse (Definition~\ref{sparsity}) it must be the case that $j=k$ and thus $\vec{v} = \vec{v_j}$. Since the threshold of every player $i$ is larger than his value in $\vec{v_j}$, except one player, it cannot be that both players $1$ and $2$ are to be allocated the item in $\vec{v_j}$.
\end{proof}

We set values for the fees $\vec{c'_i}$ and prove that $\expostmechanism'$ together with $\vec{c'_i}$ is an interim IR, dominant strategy incentive compatible and deterministic mechanism (Lemma~\ref{lemma:legal-interim}). For every active player $i \in A$ and every subset $S_j$ that he is active in we set $c'_i((v_j)_{-i}) = \frac{\frac{1-\basevectorprob{}}{\sizeActive}\cdot \feeopt{i,j}}{\basevectorprob{j}}$. For every other value of $v_{-i}$ we set $c'_i(v_{-i})=0$. 

\begin{lemma} \label{legal-fees-opt}
$\expostmechanism'$ with fees charged according to $\vec{c'_i}$ is an interim IR, dominant strategy incentive compatible and deterministic mechanism.
\end{lemma}

\begin{proof}
By Claim~\ref{legal-ex-post-lb}, $\expostmechanism'$ is a deterministic, dominant strategy incentive compatible and ex-post IR mechanism. By Proposition~\ref{characterization} it suffices to show that:
$${[CP_i(\mathcal{F})\cdot \vec{c'_i}]}_k  \leq \pi_i^{{\expostmechanism'(\mathcal{F})}}(v_{i,k}) \quad \quad  \forall i \in [n], \, \forall k=1, \dots, |\mathcal{F}_i|$$

Since the only non-zero fees are $c'_i((v_j)_{-i})$ for an active player $i \in A$ and a subset $S_j$ that he is active in, we only need to consider values $v_i \in D_i$ for which these values have non-zero probability, i.e.,  $[CP_i(\mathcal{F})\cdot \vec{c'_i}]_{(v_i,(v_j)_{-i})} > 0$. These values are $v_{i,j}, {u'}_{i,j}$ and $u_{i,j}$.

\begin{enumerate}
    \item \label{v-lb-al} $[CP_i(\mathcal{F})\cdot c'_i]_{(v_{i,j})}  \leq \pi_i^{{\expostmechanism'(\mathcal{F})}}(v_{i,j})$.
    \item \label{u'-lb-al} $[CP_i(\mathcal{F})\cdot c'_i]_{({u'}_{i,j})} \leq \pi_i^{{\expostmechanism'(\mathcal{F})}}({u'}_{i,j})$.
    \item \label{u_lb-al} $[CP_i(\mathcal{F})\cdot c'_i]_{({u}_{i,j})} \leq \pi_i^{{\expostmechanism'(\mathcal{F})}}({u}_{i,j})$.
\end{enumerate}

\begin{enumerate}
    \item \begin{align*}
         & [CP_i(\mathcal{F})\cdot c'_i]_{(v_{i,j})} =  \, \basevectorprob{i,j}\cdot c'_i((v_j)_{-i}) = \basevectorprob{i,j}\cdot \frac{\frac{1-\basevectorprob{}}{\sizeActive}\cdot \feeopt{i,j}}{\basevectorprob{j}} \leq \basevectorprob{i,j}\cdot \frac{\frac{1-\basevectorprob{}}{\sizeActive}\cdot q_{i,\sigma_i(j)}\cdot( y_{i,\sigma_i(j)} - y_{i,0})}{\basevectorprob{j}} \\
         & = \, \frac{\frac{(1-\basevectorprob{})}{\sizeActive}q_{i,\sigma_i(j)}\cdot( y_{i,\sigma_i(j)} - y_{i,0})}{\Pr\nolimits_{\mathcal{F}_i}(v_i=v_{i,j})}\\
         & =  y_{i,\sigma_i(j)} - y_{i,0} -  \frac{( y_{i,\sigma_i(j)} - y_{i,0})\cdot(\Pr\nolimits_{\mathcal{F}_i}(v_i=v_{i,j})-\frac{(1-\basevectorprob{})}{\sizeActive}q_{i,\sigma_i(j)}\cdot ( y_{i,\sigma_i(j)} - y_{i,0}))}{\Pr\nolimits_{\mathcal{F}_i}(v_i=v_{i,j})} \\
        & = \,  ( y_{i,\sigma_i(j)} - y_{i,0}) - \frac{ ( y_{i,\sigma_i(j)} - y_{i,0})(\basevectorprob{i,j} + \epsilonprob{i,j})\cdot (\Pr\nolimits_{\mathcal{F}_i}(v_i=v_{i,j}))}{\Pr\nolimits_{\mathcal{F}_i}(v_i=v_{i,j})} \\
        & = ( y_{i,\sigma_i(j)} - y_{i,0}) \cdot (1- \basevectorprob{i,j} - \epsilonprob{i,j}) \leq \pi_i^{{\expostmechanism'(\mathcal{F})}}(v_{i,j})
    \end{align*}

    \item \label{u'-al} \begin{align*}
        &[CP_i(\mathcal{F})\cdot c'_i]_{({u'}_{i,j})} =  c'_i((v_j)_{-i}) \leq  \cfrac{\frac{1-\basevectorprob{}}{\sizeActive}\cdot e}{\basevectorprob{j}} =  \cfrac{1-\basevectorprob{}}{\sizeActive}\cdot e : \cfrac{(1-\basevectorprob{})\cdot e}{\sizeActive\cdot \min\limits_{k \in \Asetforsubset{j}}{\{{u'}_{k,j}-v_{k,j} \}}} \\
        &=  \min\limits_{k \in \Asetforsubset{j}}{\{{u'}_{k,j}-v_{k,j}\}} \leq {u'}_{i,j}-v_{i,j} \leq \pi_i^{{\expostmechanism'(\mathcal{F})}}({u'}_{i,j})
    \end{align*}
    \item \begin{align*}
        & [CP_i(\mathcal{F})\cdot c'_i]_{({u}_{i,j})} =  c'_i((v_j)_{-i}) \underbrace{\leq}_{\text{Item } \ref{u'-al}} {u'}_{i,j}-v_{i,j} \leq  {u}_{i,j}-v_{i,j} \leq \pi_i^{{\expostmechanism'(\mathcal{F})}}({u}_{i,j})
    \end{align*}
\end{enumerate}

\end{proof}

The expected revenue of the interim IR mechanism ($\expostmechanism'$, $c'_i$)  with respect to $\mathcal F_S$ is:

$$ 
\sum\limits_{i=1}^{\sizeActive}{y_{i,0} \cdot \frac{1-\basevectorprob{}}{\sizeActive}} + \sum\limits_{j=1}^{m} \basevectorprob{j}\cdot \max\limits_{i \in {n}}\{v_{i,j}\} +
\sum\limits_{j=1}^m \sum\limits_{i \in \Asetforsubset{j}} \feeopt{i,j}\cdot \frac{1-\delta}{\sizeActive}
$$

\end{document}